\documentclass[useAMS,usenatbib,usegraphicx]{mn2e}
\usepackage{aas_macros}
\usepackage{amssymb}

\title[The Faint Structure of NGC 2403] {Quantifying the Faint
  Structure of Galaxies: The Late-type Spiral NGC 2403\thanks{Based on
    data collected at the Subaru telescope, which is operated by the
    National Astronomical Observatory of Japan.}\thanks{Based on
    observations made with the NASA/ESA Hubble Space Telescope,
    obtained from the Data Archive at the Space Telescope Science
    Institute, which is operated by the Association of Universities
    for Research in Astronomy, Inc., under NASA contract NAS 5-26555.
    These observations are associated with program GO10523.  }}
\author[Barker et al.]  {Michael K. Barker$^1$, Annette M. N.
  Ferguson$^1$\thanks{ferguson@roe.ac.uk}, M. J. Irwin$^2$, N. Arimoto$^{3,4}$,
  \newauthor P. Jablonka$^{5,6}$\\
  $^1$SUPA, Institute for Astronomy, University of Edinburgh, Royal Observatory, Blackford Hill, Edinburgh, UK, EH9 3HJ\\
  $^2$Institute of Astronomy, Cambridge University, Cambridge, UK, CB3 0HA\\
  $^3$National Astronomical Observatory of Japan, Mitaka, Tokyo 181-8588, Japan\\
  $^4$Department of Astronomical Science, Graduate University for Advanced Studies, Mitaka, Tokyo 181-8588, Japan\\
  $^5$Laboratoire d'Astrophysique, Ecole Polytechnique F\'ed\'erale de Lausanne (EPFL), Observatoire, CH-1290 Sauverny, Switzerland \\
  $^6$GEPI, Observatoire de Paris, CNRS UMR 8111, Universit\'e Paris
  Diderot, F-92125, Meudon, Cedex, France\\}

\begin{document}

\newcommand{\MITRGB}[0]{{\rm M_{I}(TRGB)}}
\newcommand{\vmi}[0]{\rm (F606W-F814W)}
\newcommand{\msun}[0]{\rm M_{\sun}}
\newcommand{\msunyr}[0]{\rm M_{\sun}\ yr^{-1}}
\newcommand{\msunpc}[0]{\rm M_{\sun}\ pc^{-2}}
\newcommand{\magsec}[0]{\rm mag\ arcsec^{-2}}


\date{Accepted ---- 12 September 2011. Received ----; in
  original form ----}

\pagerange{\pageref{firstpage}--\pageref{lastpage}} \pubyear{2010}

\defcitealias{Eggen62}{ELS62}
\defcitealias{Searle78}{SZ78}
\defcitealias{Pagel95}{PT95}
\defcitealias{Pagel98}{PT98}
\defcitealias{Barker07a}{Paper II}
\defcitealias{Barker07b}{B07}
\defcitealias{Williams09b}{W09b}
\defcitealias{Roskar08a}{R08a}
\defcitealias{Roskar08b}{R08b}

\maketitle

\label{firstpage}

\begin{abstract} 

  Ground-based surveys have mapped the stellar outskirts of Local
  Group disc galaxies in unprecedented detail, but extending this work
  to other galaxies is necessary in order to overcome stochastic
  variations in evolutionary history and provide more stringent
  constraints on cosmological galaxy formation models.  As part of our
  continuing program of ultra-deep imagery of galaxies beyond the
  Local Group, we present a wide-field analysis of the isolated
  late-type spiral NGC~2403 using data obtained with Suprime-Cam on
  the Subaru telescope.  The surveyed area reaches a maximum projected
  radius of 30 kpc or deprojected radius of $R_{dp} \sim 60$ kpc.  The
  colour-magnitude diagram reaches 1.5 mag below the tip of the
  metal-poor red giant branch (RGB) at a completeness rate $> 50\%$
  for $R_{dp} \gtrsim 12$ kpc.  Using the combination of diffuse light
  photometry and resolved star counts, we are able to trace the radial
  surface brightness (SB) profile over a much larger range of radii
  and surface brightness than is possible with either technique alone.
  The exponential disc as traced by RGB stars dominates the SB profile
  out to $\ga 8$ disc scale-lengths, or $R_{dp} \sim 18$ kpc, and
  reaches a $V$-band SB of $\mu_V \sim 29\ \magsec$.  Beyond this
  radius, we find evidence for an extended structural component with a
  significantly flatter SB profile than the inner disc and which we
  trace to $R_{dp} \sim 40$ kpc and $\mu_V \sim 32\ \magsec$.  This
  component can be fit with a power-law index of $\gamma\sim3$, has an
  axial ratio consistent with that of the inner disc and has a V-band
  luminosity integrated over all radii of 1--7\% that of the whole
  galaxy.  At $R_{dp} \sim 20 - 30$ kpc, we estimate a peak
  metallicity [M/H]~$= -1.0 \pm 0.3$ assuming an age of 10 Gyr and
  zero $\alpha-$element enhancement.  Although the extant data are
  unable to discriminate between stellar halo or thick disc
  interpretations of this component, our results support the notion
  that faint, extended stellar structures are a common feature of all
  disc galaxies, even isolated, low-mass systems.

\end{abstract}


\section{Introduction}
\label{sec:intro}

The stellar outskirts of galaxies are important testing grounds for
models of galaxy formation and evolution.  This is because the
dynamical and star formation timescales there are relatively long
making it easier to identify accreted material and to study relatively
unprocessed gas.  N-body and hydrodynamical simulations of galaxy
formation within a cosmological context predict that the merging and
accretion that is more common at high redshift can leave an imprint on
galaxy outskirts that is visible to the present day in the form of
thick discs, stellar haloes and discrete substructures
\citep[e.g.][]{Brook04,Bullock05,Cooper10}.  Within this scenario, the
properties of these structures may correlate with host galaxy
properties, like present-day total mass, but they are also expected to
exhibit significant variations at the same mass scale due to
stochastic variations in the merging/accretion history and the
detailed nature of the individual progenitor systems
\citep[e.g.][]{Cooper10,Purcell07}.  Thick discs and haloes may also
arise from other processes besides merging, such as radial migration,
misaligned gas accretion, and in-situ star formation
\citep{Schonrich09,Roskar10,Loebman10}.  Therefore, it is crucial to
study as many galaxies and galaxy types as possible to overcome
stochastic variations and discern underlying trends that may help to
isolate the dominant formation mechanisms.

These outer stellar structures are very faint, typically several
magnitudes below the sky level.  Detecting their diffuse light
requires very careful treatment of sky subtraction, flat-fielding
errors, detector response, scattered light, and PSF wings
\citep[e.g.][]{Morrison97,DeJong08b}.  Nevertheless, a growing body of
diffuse light analyses supports the idea that such structures are
common around disc galaxies
\citep[e.g.][]{Zibetti04,Zibetti04b,Jablonka10,Burstein79,Tsikoudi79,
  Shaw90,deGrijs96,Morrison97,Neeser02,Dalcanton02,
  Malin97,Shang98,MartinezDelgado09}.  Most of these studies have
imaged in a single band hence there is no information on the the
nature of the extended stellar populations. Even in cases where
multiple passbands have been obtained, the age-metallicity degeneracy
present in optical broadband colours enables only very crude
constraints.

An alternative approach to studying galaxy outskirts is with resolved
stars, a technique which can typically reach far fainter surface
brightness (SB) levels than diffuse light.  The most interesting
cosmological constraints come from the old stars in these systems,
those on the red giant branch (RGB).  With ground-based telescopes,
resolving RGB stars in external galaxies was initially limited to
systems within the Local Group
\citep[e.g.][]{Ferguson02,Ferguson07a,Ibata07, Mcconnachie09}.  These
studies found a wealth of very faint stellar structures around the
MW-analog M31 and comparatively little around the late-type spiral
M33.  These structures exhibited large-scale inhomogeneities in
distribution and composition, highlighting the importance of areal
coverage when looking at galaxy outskirts.  However, these are just
two systems; more rigorous tests of cosmological galaxy formation
models require similar data for many more galaxies beyond the Local
Group.

With this motivation in mind, we are conducting a program to explore
the low surface brightness outer regions of all large galaxies within
5 Mpc using wide-field imagers on 8-m class telescopes. In our first
paper, we used Subaru/Suprime-Cam to identify an extended structure of
RGB stars around the MW-analog, M81, stretching out to a deprojected
radius $R_{dp} = 44$ kpc \citep{Barker09}.  This structure had a
flatter radial and azimuthal surface density profile than the main
disc suggesting it was a halo or thick disc, but its properties did
not exactly match either of these components in the MW.  Furthermore,
as M81 is part of an interacting group of galaxies, we could not
exclude the hypothesis that the extended component was the result of a
recent tidal encounter.

In parallel with our efforts, other groups have pursued similar
studies but typically with smaller field-of-view (FOV) imagers
\citep[e.g.][]{ Vlajic09, Vlajic11,deJong08,Rejkuba09}.  Such studies
risk being affected by the presence of localized substructures which
fall within the FOV and also suffer from significant uncertainty in
the background/contaminant subtraction which is a crucial aspect of
quantifying low surface brightness emission.  Even for systems beyond
the Local Group, the FOVs of HST and GMOS are too small to reveal a
{\it global} picture of their outer structures and wide-field imagers
like Suprime-Cam are clearly needed \citep[e.g.][]{Mouhcine10,
  Bailin11,Tanaka2011}.  Perhaps it is not so surprising then, that,
until recently, no detailed {\it global analysis} of RGB stars had
been conducted for a spiral galaxy outside the Local Group.

As part of our program, we present here observations of the late-type
Sc galaxy, NGC~2403.  With a total mass $\sim 10^{11} M_{\sun}$ and a
circular velocity of $\sim 135\ \rm km\ s^{-1}$ \citep{Fraternali02},
NGC~2403 is similar in many respects to M33 and NGC 300, and is,
therefore, a good system with which to increase the observed baseline
in galaxy mass.  In this work, we adopt the HST Key Project
Cepheid-based distance modulus of $(m-M)_0 = 27.48 \pm 0.24$, giving
it a distance of 3.13 Mpc \citep{Freedman01}.  This distance compares
favorably to HST RGB tip-based distances of 3.09 -- 3.20 Mpc
\citep{Dalcanton09}.  At this distance, $1.0\arcmin \approx 0.9$ kpc.
NGC 2403 is located $\sim 30\degr$ from the plane of the MW with a
Galactic longitude of $\sim 150\degr$.  It has a B-band isophotal
radius of $10.9\arcmin$, or 9.8 kpc, and a photographic V-band disc
scale-length $h = 1.5$ kpc within the inner 4.5 kpc \citep{Okamura77}.
The total apparent V-band magnitude, corrected for reddening, is $m_V
= 8.04$ \citep{deVauc91}, which translates to an absolute magnitude of
$M_V = -19.44$.  With an inclination of $63\degr$
\citep{Fraternali02}, inclination-dependent extinction effects are not
likely to be significant ($\sim 0.3$~mag, \citet{Shao07}).

\citet{Fraternali02} used high resolution Very Large Array
observations of neutral hydrogen in NGC~2403 to derive a total HI mass
$\sim 3 \times 10^8 M_{\sun}$ and a dynamical mass $\sim 10^{11}
M_{\sun}$ inside a radius of 23 kpc.  These observations revealed no
signs of interaction in the immediate vicinity.  However, they did
reveal an asymmetric warp in the outer disc and a thick, clumpy layer
of HI that rotates more slowly than, and contains roughly $10\%$ the
mass of, the cold HI disc.

NGC~2403 is the brightest member of a loose galaxy group that shows no
clear signs of interactions.  \citet{Chynoweth09} detected no HI
clouds in this group with masses down to a limit of $2.2 \times 10^6
M_{\sun}$.  The nearest systems to NGC 2403 are the small galaxies,
DDO 44 and NGC 2366, each lying $\sim 80$ kpc and $\sim 200$ kpc away
in projection, respectively.  Five other small galaxies are located
$\sim 350$ kpc away in projection \citep{Karachentsev02}.  On larger
scales, NGC~2403 belongs to a filament of $\sim 60$ known galaxies
spanning roughly $25\degr$ on the sky.  M81 is located near the center
of the filament $\sim 800$ kpc from the spirals NGC~2403 and NGC 4236,
which lie at opposite ends.  NGC 2403 is at least 4 times farther away
from the nearest large disc galaxy (M81) than M33 is from M31 and
hence it can be considered a far more isolated system.

This paper is organized as follows.  In \S \ref{sec:obs}, we describe
the observations and data reduction procedures.  In \S \ref{sec:cmd},
we present the colour-magnitude diagram (CMD) and, in \S
\ref{sec:spatial}, we show the 2-dimentional point source maps.  Then,
the radial star count profiles are derived in \S \ref{sec:sdprof},
which reveal the presence of the extended component.  After that, in
\S \ref{sec:mdf}, we estimate the RGB metallicity of the extended
component.  In \S \ref{sec:sbprof}, the SB profile is derived and we
fit the SB profile with several models and, in \S \ref{sec:disc}, we
discuss the implications.  Finally, we summarize the results in \S
\ref{sec:summ}.

\section{Observations and Data Reduction}
\label{sec:obs}

\begin{figure}
\includegraphics[width=80mm,height=150mm,keepaspectratio=true]{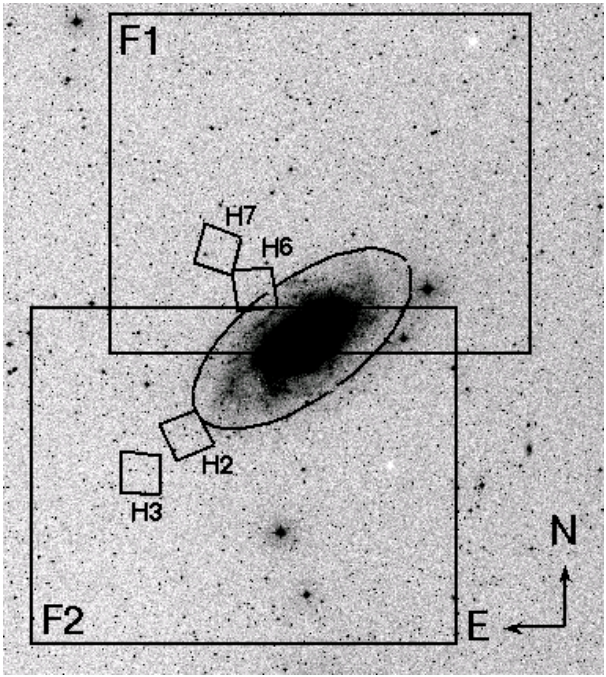}
\caption{Digitized Sky Survey image showing the sizes and locations of
  the two fields observed with Suprime-Cam (F1 and F2).  The ellipse
  marks the $R_{25}$ radius of $10.9\arcmin$ or $9.8$ kpc.  Small
  boxes show the four ACS fields discussed in \S \ref{sec:mdf}.}
\label{fig:fields}
\end{figure}

The observations were obtained with the Suprime-Cam instrument
\citep{Miyazaki02} on the 8-m Subaru telescope on the night of January
8, 2005 (S04B, PI=N. Arimoto).  This instrument consists of 10 CCDs of
2048x4096 pixels arranged in a 2x5 pattern, with a pixel scale of
0.2~arcsec and a total field of view of approximately 34\arcmin
x27\arcmin (including long edge inter-chip gaps of 16 -- 17 arcsec and
short edge gaps of 5 -- 6 arcsec).

\begin{figure*}
\includegraphics[width=6in,keepaspectratio=true]{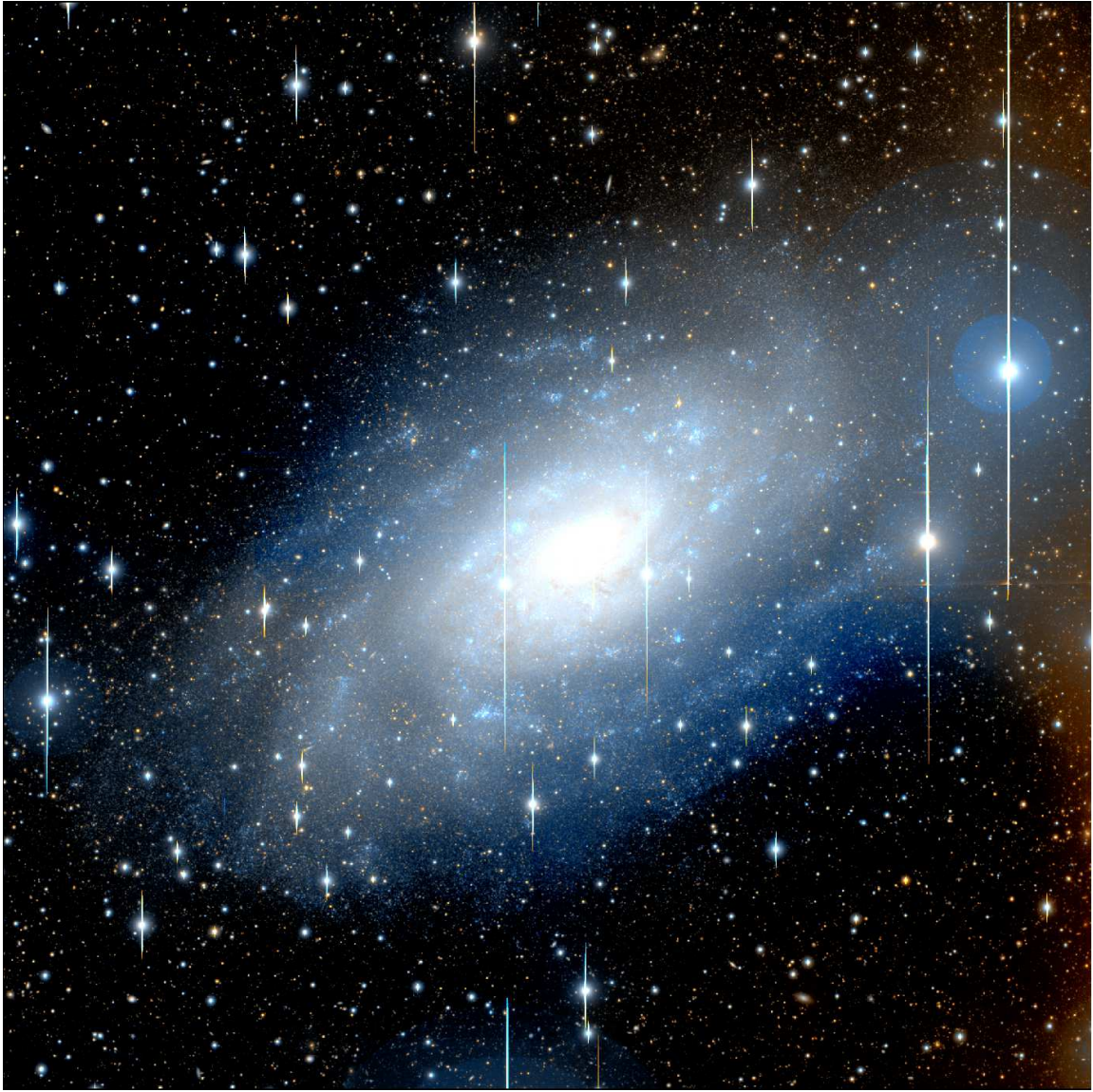}
\caption{Colour composite image of NGC~2403 created from the stacked
  mosaics and cropped to $\approx 24$ kpc on a side.  The intensity
  scaling is logarithmic and the colour mapping is similar to that of
  \citet{Lupton04} with $V$ for the blue channel, $i'$ for red, and
  the average for green.  Luminous main sequence and evolved giant
  stars can be seen to resolve in the outer parts of the disc.  North
  is up and east is to the left.}
\label{fig:mosaic}
\end{figure*}

NGC~2403 was covered using two field centers, one to the north of its
nucleus at $(\alpha_{J2000},\delta_{J2000}) = (7^h 36^m
36^s.7,+65^\circ 49' 26'')$ and another to the south at
$(\alpha_{J2000},\delta_{J2000}) = (7^h 37^m 38^s.7,+65^\circ 24'
31'')$.  The large rectangular boxes in Fig.~\ref{fig:fields} outline
the locations of these fields, which we refer to as F1 and F2,
respectively.  For each field, we obtained a set of 8 images in the
Johnson $V$ filter with individual exposure times of 450s, and 12
images in the Sloan $i'$ filter with exposure times of 205s. All
observations were recorded under slightly non-photometric conditions
through patchy thin cirrus.  The images of F1 were taken in an average
seeing of $\sim 0.8\arcsec$ in both filters while the F2 images had an
average seeing of $\sim 1.1\arcsec$ and $\sim 0.9\arcsec$ in the
$V$-band and $i'$-band, respectively.

To fill in the chip gaps and facilitate the removal of cosmic rays and
bad pixels, individual images were dithered by $\sim 25$ arcsec,
resulting in a mosaic for each field covering $\approx$ 36\arcmin
x28\arcmin, or $\approx$ 32x25 kpc.  Accounting for the overlap region
between F1 and F2, the total surveyed area was $\approx 0.54\rm \
deg^{2}$, or $\approx 1800\rm \ kpc^{2}$, in the central 39x48 kpc
around NGC 2403.  Flat-field and interchip gain variations were
removed with master flats obtained by combining 12 and 11 twilight sky
flats in the $V$ and $i'$ filters, respectively. After flat-fielding,
remaining large scale variations in dark sky level, measured directly
from stacked dark sky images at several different positions obtained
during this run, were less than 1\% of sky.  An $i'$-band fringe frame
acquired from an earlier Suprime-Cam imaging run was used to help
assess the degree of dark sky fringing present, but this was found to
be negligible in our data, so this extra image processing step was not
required.

The image processing procedure was very similar to that followed by
\citet{Barker09}.  After converting the raw data to multi-extension
FITS format, all images and calibration frames were run through a
variant of the data reduction pipeline developed for the Isaac Newton
Telescope (INT) Wide Field Survey (WFS)
\footnote{http://www.ast.cam.ac.uk/$\sim$wfcsur/}.  Here we present a
brief overview of the main steps of the pipeline which is described in
more detail in \citet{Irwin85,Irwin97,Irwin01} and \citet{Irwin04}.

Prior to deep stacking, catalogues were generated for each individual
processed science image to both refine the astrometric calibration and
asses the data quality.  For astrometric calibration, a Zenithal
polynomial projection \citep{Greisen02} provided a good prescription
for the World Coordinate System (WCS) and included all the significant
telescope radial field distortions. We used this in conjunction with a
6-parameter linear plate model per detector to define the required
astrometric transformations.  The 2MASS point source catalogue
\citep{Cutri03} was used for the astrometric reference system.

The individual image qualities were then assessed using the average
seeing and ellipticity of stellar images, as well as sky level and sky
noise determined from the object cataloguing stage.  Images were
stacked at the detector level using the updated WCS information to
accurately align them to a reference image.  The background level in
the overlap area between each image in the stack and the reference was
adjusted additively to compensate for sky variations during the
exposure sequence and the final stack included seeing weighting,
confidence (i.e. variance) map weighting, and clipping of cosmic rays.

Next, we generated detector-level catalogues from the stacked images
and updated the WCS astrometry in the FITS image extensions prior to
mosaicing all detectors together.  Residual astrometric errors over
the whole stacked array were typically $< 0.1\arcsec$, greatly
simplifying this process.  Slight offsets in underlying sky level
between the stacked detector images caused small (typically $\sim
0.1-0.2\%$ of sky), but still visible, discontinuities in the final
mosaics.  These offsets were due to small colour equation differences
in the detectors and the relatively blue colour of the twilight sky
compared to dark sky and unresolved diffuse light from NGC~2403.  We
corrected these offsets iteratively by visual inspection of a 4x4
blocked version of the mosaics using a pre-assigned keyword in each
relevant detector FITS extension designed for this purpose.

Fig.~\ref{fig:mosaic} shows a colour composite image made from the $V$
and $i'$ stacked mosaics of both fields and cropped to $\approx 24$
kpc on a side.  In this image, north is up and east is to the left.
The intensity scaling is logarithmic and the colour mapping is similar
to that of \citet{Lupton04} with $V$ for the blue channel, $i'$ for
red, and the average for green.

The data reduction pipeline also produced an aperture photometry
catalogue using a ``soft-edged'' aperture with radius close to the
full-width-half-maximum (FWHM) \citep[e.g.,][]{Irwin97,Naylor98}.  A
series of apertures ranging from 1/2 to 4 times the FWHM were
additionally used to compute stellar aperture corrections and to
correct for PSF variation across the field-of-view.

The photometric calibration was based on a comparison with 0.29
deg$^2$ of INT $V, i'$-band photometry centred on NGC 2403 taken in
photometric conditions during April 2009.  The INT photometry were
converted to the Johnson-Cousins $V, I$ system using the
transformation equations given on the WFS website.  All the INT data
were calibrated on a nightly basis against multiple observations of
Landolt standard stars.  Both Subaru fields for NGC 2403 were compared
independently with the INT data and were found to have the same
zero-points to within their errors ($1-2\%$).  They were also checked
directly against each other using the common overlap region and again
they agreed within this error.  Thus, the photometric zero-point is
accurate to $\sim 0.02$~mag.

\subsection{PSF-Fitting Photometry}
\label{sec:phot}

\begin{figure}
\includegraphics[width=3in,keepaspectratio=true]{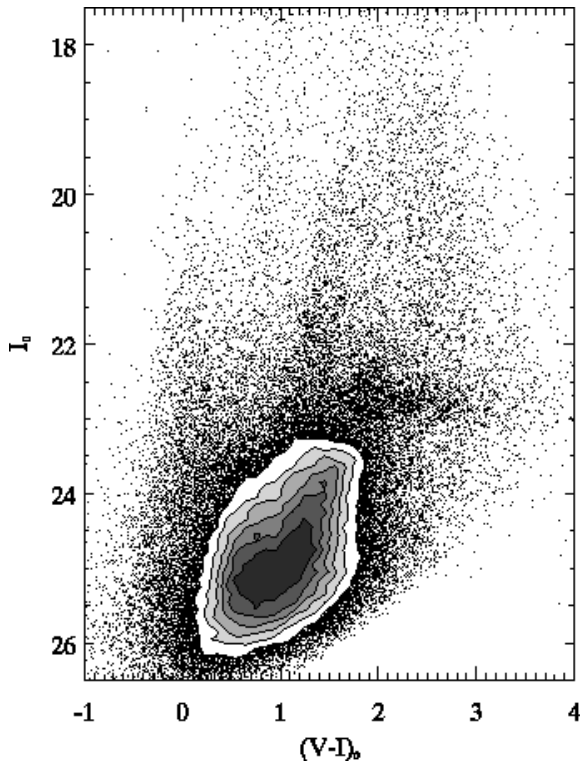}
\caption{CMD of all point sources in F1 and F2.  The contours indicate
  [140, 200, 260, 320, 380, 440] $\rm stars\ decimag^{-2}$.}
\label{fig:cmd}
\end{figure}

Because the data reduction pipeline did not provide a direct estimate
of the completeness rate, we elected to additionally perform
PSF-fitting photometry on the stacked mosaics.  This was accomplished
using the standalone versions of the DAOPHOT/ALLSTAR/ALLFRAME suite of
programs \citep{Stetson87,Stetson88,Stetson94}.  The PSF for each
field in each filter was built by starting with an initial list of
roughly 1000 bright, fairly isolated stars and iteratively subtracting
neighbors, rejecting stars with large residuals, and increasing the
spatial complexity of the PSF as a function of position on the
mosaics.  In the end, this left several hundred stars to build a PSF
that varied quadratically with position on each stacked mosaic.

 \begin{figure}
  \includegraphics[width=3in,keepaspectratio=true]{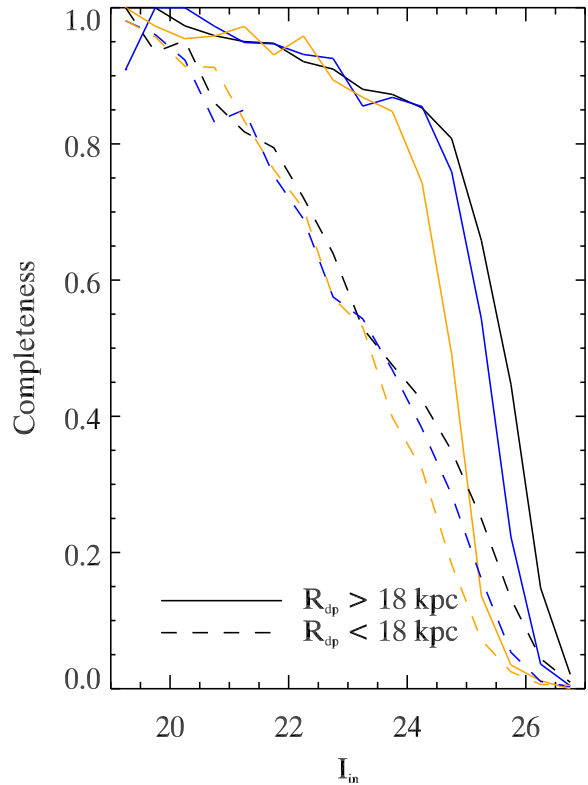}
  \caption{ Completeness rate as a function of input magnitude as
    derived from artificial star tests.  The curves show the
    completeness for two radial ranges, $R_{dp} < 18$ kpc (dashed) and
    $R_{dp} > 18$ kpc (solid) in both fields combined, and for three
    different input (V$-$I) colour ranges, 0 -- 1 (black), 1 -- 2
    (blue), and 2 -- 3 (orange). }
\label{fig:comp}
\end{figure}

We derived a coordinate transformation between the stacked $V$ and
$i'$ mosaics of each field using DAOMASTER \citep{Stetson93}.  The
stacked mosaics were then coadded with MONTAGE2 \citep{Stetson94}.
Objects meeting a $3\sigma$ detection threshold on the coadded image
were measured with ALLSTAR to obtain a first guess at positions and
magnitudes and the resulting list was input into ALLFRAME together
with the stacked mosaics in each filter and their coordinate
transformation.  To reduce contamination from non-stellar sources in
the ALLFRAME photometric catalogue, we excluded objects with
abnormally low or high sharp values as measured in the $i'$-band.  For
the overlap region between F1 and F2 we used only the F1 catalogue
since it had better seeing.  The PSF magnitudes were standardized to
the Johnson-Cousins system using linear transformations to the
calibrated $V, I$ Subaru aperture photometry described in \S
\ref{sec:obs}.

Fig.~\ref{fig:cmd} shows the colour-magnitude diagram (CMD) for the
final point source catalogue for both fields, which contains 163293
sources.  Contours are overplotted where the density of points is
extremely high.  The contour levels correspond to [140, 200, 260, 320,
380, 440] $\rm stars\ decimag^{-2}$.  There are several clear stellar
sequences visible, which we discuss in more detail in \S
\ref{sec:cmd}.

\subsection{Artificial Star Tests}
\label{sec:fake}

\begin{figure}
\includegraphics[width=3in,keepaspectratio=true]{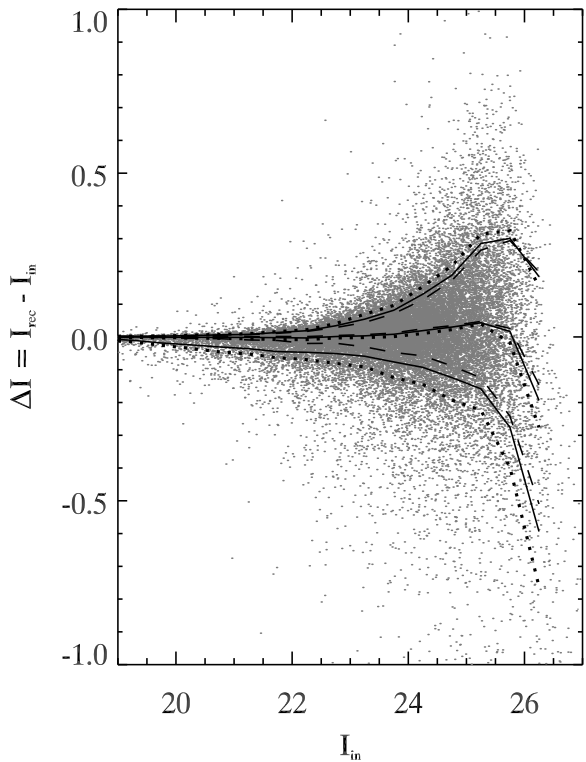}
\caption{ The difference between the recovered and input $I$-band
  magnitudes of the artificial stars in both fields combined.  The
  lines show the median shifts and the central 68\% for all stars
  (solid) and those with $R_{dp} < 18$ kpc (dotted) and with $R_{dp} >
  18$ kpc (dashed).  There are no significant systematic magnitude
  shifts for $I \lesssim 25.5$.  }
\label{fig:magerr}
\end{figure}

\begin{figure}
\includegraphics[width=3in,keepaspectratio=true]{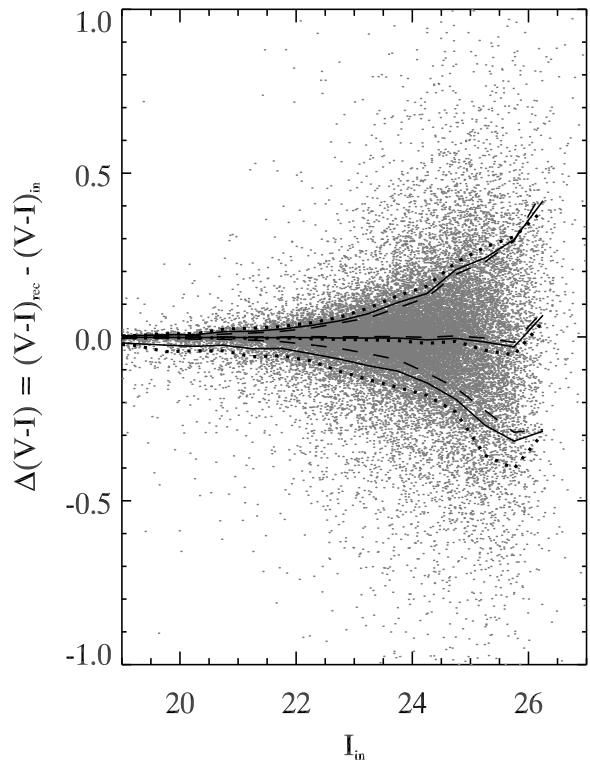}
\caption{ The difference between the recovered and input colours of
  the artificial stars in both fields combined.  The lines show the
  median shifts and the central 68\% for all stars (solid) and those
  with $R_{dp} < 18$ kpc (dotted) and with $R_{dp} > 18$ kpc (dashed).
  There are no significant systematic colour shifts for $I \lesssim
  25.5$.  }
\label{fig:colerr}
\end{figure}

To estimate completeness, we performed artificial star tests in which
$\sim 12000$ fake stars with known magnitudes were injected onto each
stacked mosaic and then the PSF-fitting procedure described above was
repeated.  The fake stars had positions, magnitudes, and colours
distributed like the real stars.  This process was repeated several
times to build up a total catalogue of $\sim 140000$ injected fake
stars equally divided between F1 and F2.

Fig.\ \ref{fig:comp} plots the completeness rate derived from the
artificial star tests for both fields combined.  Close to the bright,
optical disc of NGC~2403, the completeness varies significantly with
deprojected radius ($R_{dp}$) \footnote{Deprojected radii refer to the
  circular radii within the disc plane and are calculated assuming the
  2MASS near-infrared isophotal center of
  $(\alpha_{J2000},\delta_{J2000}) = (7^h 36^m 51^s.4,+65^\circ 36'
  9'')$ \citep{Jarrett03}, an inclination of 63\degr, and a position
  angle of 124\degr\ measured north through east
  \citep{Fraternali02}.}.  Thus, we show the completeness for two
radial ranges, $R_{dp} < 18$ kpc (dashed) and $R_{dp} > 18$ kpc
(solid), and for three different input (V$-$I) colour ranges, 0 -- 1
(black), 1 -- 2 (blue), and 2 -- 3 (orange).  Table 1 lists the 50\%
completeness levels in the $I$-band for both fields individually and
for the total catalogue.  In F1, the 50\% completeness level occurs at
$\sim 23.5 - 24.2$ for $R_{dp} < 18$ kpc and at $\sim 24.9 - 25.9$ for
$R_{dp} > 18$ kpc.  In F2, the 50\% completeness level occurs at $\sim
22.9 - 23.1$ for $R_{dp} < 18$ kpc and at $\sim 24.4 - 25.3$ for
$R_{dp} > 18$ kpc.  The 50\% level is fainter in F1 than it is in F2
because of the difference in seeing, but the effect is most noticeable
in the bright optical disc where the crowding is highest.  We account
for this difference whenever necessary by treating the completeness
corrections separately for each field.  In \S \ref{sec:sdprof}, we
further examine the completeness as a function of radius in the
context of the star count profiles.

\begin{table}
 \centering
 \begin{minipage}{3in}
  \caption{$I$-band $50\%$ completeness levels.}
  \begin{tabular}{lccc}
  \hline
  $(V-I) =$    &   0 -- 1   & 1 -- 2     &  2 -- 3   \\
 \hline
  & &  $R_{dp} < 18$ kpc & \\
 \hline
Field F1           & 24.2 & 23.9 & 23.5 \\
Field F2           & 23.1 & 23.2 & 22.9 \\
Total catalogue        & 23.5 & 23.5 & 23.4 \\
 \hline
  & &  $R_{dp} > 18$ kpc & \\
 \hline
Field F1           & 25.9 & 25.5 & 24.9 \\
Field F2           & 25.3 & 25.0 & 24.4 \\
Total catalogue        & 25.6 & 25.3 & 24.7 \\
\hline
\end{tabular}
\end{minipage}
\label{tab:50comp}
\end{table}

In Fig.\ \ref{fig:magerr}, we show the $I$-band photometric shifts of
the artificial stars, $\Delta I$, for the total catalogue.  This shift
is defined as the recovered magnitude minus the input magnitude.  The
lines show the median shift and the central 68\% for stars at all
radii (solid) and those with $R_{dp} < 18$ kpc (dotted) and with
$R_{dp} > 18$ kpc (dashed).  For all radii, the median of $| \Delta I
|$ is $\approx 0.1$ at $I = 24.7$ and $\approx 0.2$ at $I = 25.9$.
Fig.\ \ref{fig:colerr} plots the same information as Fig.\
\ref{fig:magerr}, but for the colour shift, $\Delta (V-I)$.  For all
radii, the median of $| \Delta (V-I) |$ is $\approx 0.1$ at $I = 24.5$
and $\approx 0.2$ at $I = 25.7$.  These figures show that there are no
significant systematic colour or magnitude shifts for $I \lesssim
25.5$.

\section{Colour-Magnitude Diagram}
\label{sec:cmd}

\begin{figure}
\includegraphics[width=3in,keepaspectratio=true]{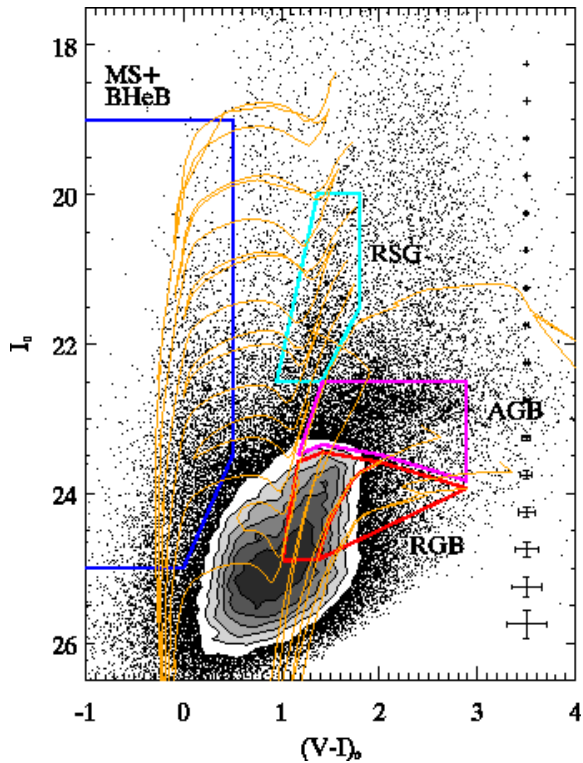}
\caption{Point source CMD with theoretical isochrones from
  \citet{Marigo08} overlaid.  The young isochrones have ages of 10.0,
  17.8, 31.6, 56.2, 100, and 178 Myr and a metallicity [M/H] $= -0.4$.
  The three old isochrones have a common age of 10 Gyr and [M/H] =
  --1.3, --0.7, and --0.4.  The boxes are used to select stars in
  different evolutionary stages: main sequence and blue helium burning
  (MS+BHeB), red supergiant (RSG), asymptotic giant branch (AGB), and
  red giant branch (RGB).  The error bars on the right-hand side show
  typical photometric errors derived from artificial star tests.  }
\label{fig:cmdiso}
\end{figure}

In Fig.~\ref{fig:cmdiso}, we show the de-reddened point source CMD
with isochrones from \citet{Marigo08} shifted to the distance of NGC
2403.  On the right-hand side are typical photometric errors (the
median of $|\Delta I|$) from the artificial star tests.  Clearly,
there is a range of ages and metallicities present in these fields.
The young isochrones at $(V-I)_0 \sim 0$ have ages of 10.0, 17.8,
31.6, 56.2, 100, and 178 Myr and a metallicity [M/H]~$\approx{\rm
  log}(Z/Z_{\sun}) = -0.4$.  This metallicity should be representative
of the young populations in the disc, as \citet{Garnett97} measured
[O/H] of HII regions to decrease from roughly solar at 1 kpc to about
0.4 dex below solar at 6 kpc.  The three old isochrones at $(V-I)_0
\sim 1 - 3$ have a common age of 10 Gyr and [M/H] = --1.3, --0.7,
--0.4.  The discontinuities in the asymptotic giant branch (AGB) are
explained in \citet{Marigo08} and are caused by changes in the opacity
tables at the transition to the thermally pulsing phase.

\begin{figure}
\includegraphics[width=3in,keepaspectratio=true]{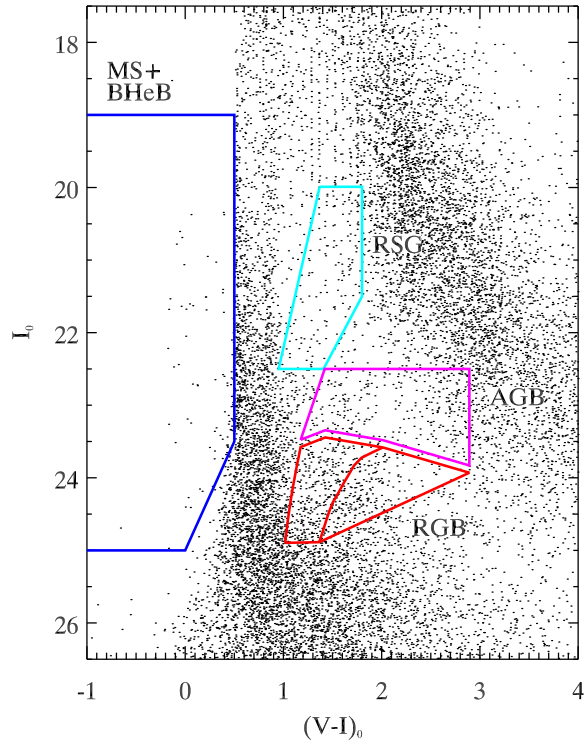}
\caption{ CMD of foreground stars based on the Besan\c{c}on model of
  the Milky Way \citep{Robin03} for a field with the same location and
  total area as F1 and F2.  The stars' magnitudes were scattered
  according to the artificial star tests, but no completeness
  corrections were applied.  The number of predicted foreground stars
  is $\sim 8\%$ of the number of point sources over the observed
  magnitude range.  The selection boxes avoid the most heavily
  contaminated regions.  }
\label{fig:cmdfground}
\end{figure}

We apply extinction corrections on a star-by-star basis using the
\citet{Schlegel98} maps.  These maps indicate a median $E(B-V)$ value
of 0.04 with negligible spatial variation.  We adopt the
\citet{Cardelli89} extinction law, for which $R_V = 3.1$ and $A_I/A_V
= 0.479$.  The inner $\sim 10$ kpc in projected radius around NGC~2403
were masked in the \citet{Schlegel98} maps and replaced with median
values from the surrounding sky.  Thus, the star-by-star correction
does not include extinction internal to NGC~2403, but this should not
be a serious problem since (i) we mainly focus on regions outside the
bright optical disc, (ii) we are mostly concerned with the RGB stars,
which tend to lie farther away from the high extinction star forming
regions than young stars \citep{Zaritsky99}, and (iii) our CMD
selection boxes are large compared to the expected amount of internal
extinction.

\begin{figure}
\includegraphics[width=3in,keepaspectratio=true]{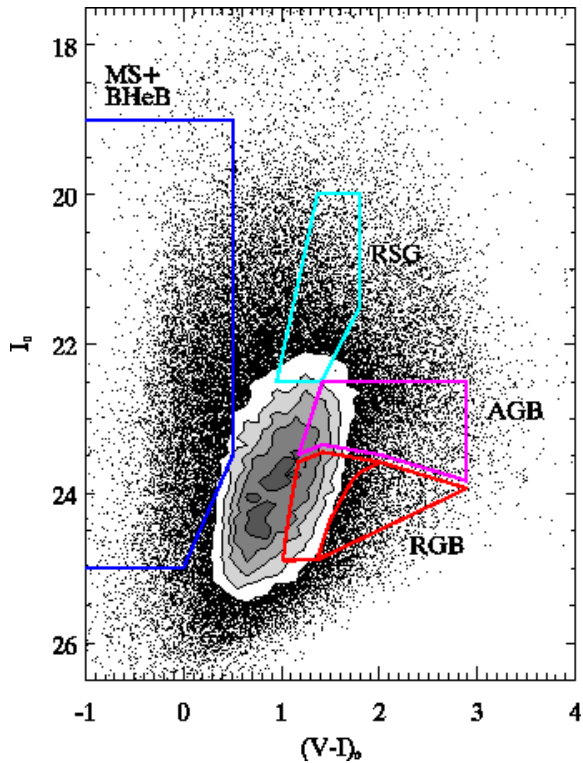}
\caption{ CMD of extended objects.  The selection boxes avoid the most
  heavily contaminated region.  The total number of extended objects
  is approximately equal to the number of point sources.  }
\label{fig:cmdp1}
\end{figure}

In what follows, we will focus on several particular CMD regions
(outlined in Fig.~\ref{fig:cmdiso}) which isolate stars in different
evolutionary stages in NGC~2403.  The blue lines in
Fig.~\ref{fig:cmdiso} mark the region occupied by young main sequence
and blue helium burning stars (MS+BHeB) with ages $\sim 10 - 150$ Myr.
Stars within the cyan polygon are red supergiants (RSGs) with ages in
the range $\sim 20 - 180$ Myr.  The red lines enclose RGB stars, which
can have ages $\sim 1 - 10$ Gyr.  The bright edge of the RGB box is
set by the RGB tip of the isochrones which has a small metallicity
dependence.  There could be some contamination of the RGB box by AGB
stars and by young, red helium burning stars with masses of $\sim 3-4\
M_{\sun}$, particularly if they have [M/H] $> -0.4$.  The magenta
lines enclose AGB stars above the RGB tip, which tend to have somewhat
younger ages ($\sim 0.5 - 8$ Gyr) than the RGB stars
\citep{MartinezDelgado99,Gallart05}.

We have also divided the RGB box into ``metal-poor'' ([M/H] $\lesssim
-0.7$) and ``metal-rich'' ([M/H] $\gtrsim -0.7$) subregions. There
will be some overlap in the metallicities probed by these subregions
due to photometric errors, but they are broader than the errors, and
so are useful in identifying any population gradients.  Some of the
most metal-poor ([M/H] $\lesssim -2.3$) and metal-rich ([M/H] $\gtrsim
-0.4$) RGB stars may fall outside the total RGB box, but extending it
further to the blue or red would increase contamination from MW
foreground stars and unresolved background galaxies, and would
increase uncertainties due to incompleteness.  We note that our use of
the term, metal-poor, differs somewhat from the traditional sense
because it includes metallicities up to [M/H] $= -0.7$ but this
definition has been chosen for consistency with \citet{Barker09}.

There is a clear RGB sequence visible in Fig.\ \ref{fig:cmdiso} with a
colour distribution that peaks between the [M/H] = --1.3 and --0.7
isochrones, suggesting a dominant population in between these two
metallicities.  At magnitudes fainter than $I_0 \sim 25$, the peak in
the colour distribution moves toward bluer colours because of the
increasing contamination from unresolved background galaxies.  To
mitigate this contamination, we do not use any sources fainter than
$I_0 = 25$.

Fig.~\ref{fig:cmdfground} shows the foreground star CMD predicted by
the Besan\c{c}on model of the MW \citep{Robin03} for the same total
area and line of sight as our observations.  We applied extinction
corrections to the foreground stars in the same way as for the real
data.  The stellar colours and magnitudes have been scattered using a
simple exponential function to mimic the increase of photometric error
with magnitude seen in the artificial star tests.  The Besan\c{c}on
model predicts that the number of foreground stars is $\sim 8\%$ of
all point sources over the magnitude limits of the NGC~2403 CMD.
There are two main features in the foreground star CMD, a narrow
vertical strip at $(V-I)_0 \sim 0.5 - 1.0$ and a broader, curved band
in the upper right quadrant.  The vertical strip is composed mostly of
main sequence turnoff stars in the MW halo at $I_0 \gtrsim 20$ and in
the MW disc at $I_0 \lesssim 20$ while the curved band is comprised of
late-type main sequence stars in the MW disc.  These two features are
visible in the NGC~2403 CMD at magnitudes brighter than $I_0 \sim 22$.
As can be seen in Fig.~\ref{fig:cmdfground}, the CMD selection boxes
sample regions that minimize contamination from foreground stars.

To check the effectiveness of our morphological classification,
Fig.~\ref{fig:cmdp1} shows the CMD of the objects classified as
extended (i.e., with high sharp values indicating a poor fit to the
stellar PSF).  The CMD selection boxes and contours are overlaid to
facilitate comparison with Fig.~\ref{fig:cmdiso}.  The number of
extended objects is about the same as the number of point sources.
The extended objects are concentrated in a broad diagonal band, the
bulk of which lies at bluer colours than the RGB and AGB selection
boxes.  Importantly, the extended objects have a colour-magnitude
distribution that is different from that of the point sources, and
most of the stellar sequences in Fig.~\ref{fig:cmdiso} are not
visible.  Examination of the extended object spatial distribution in
the sky reveals that some of them are misclassified stars located in
heavily crowded regions in the bright optical disc of NGC~2403.
However, these are the most poorly measured objects and they should
not affect our conclusions because we apply completeness corrections
to the radial star count profiles and we rely on the diffuse light
profile inside $R_{dp} \sim 9$ kpc.

\section{Star Count Analyses}

\subsection{Spatial Distribution}
\label{sec:spatial}

\begin{figure*}
\includegraphics[width=6in,keepaspectratio=true]{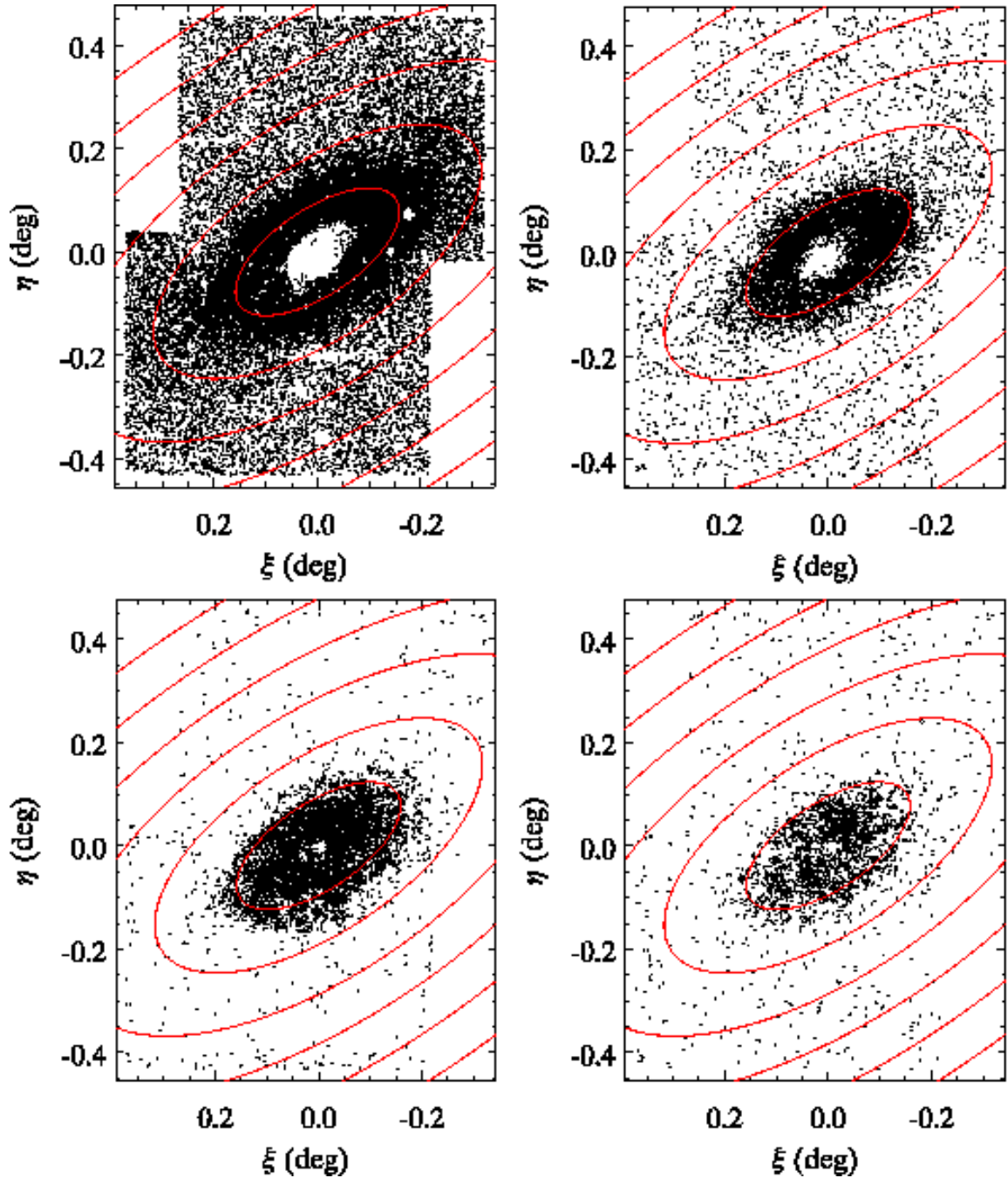}
\caption{ Going clockwise from top left, the tangent plane projection
  of RGB stars (ages $\sim 1 - 10$ Gyr), AGB stars (ages $\sim 0.5 -
  8$ Gyr), RSG stars (ages $\sim 20 - 180$ Myr), and MS+BHeB stars
  (ages $\sim 10 - 150$ Myr).  Ellipses denote deprojected radii of 10
  -- 60 kpc in steps of 10 kpc (1 kpc $\approx 1.1\arcmin$).  No
  correction for contaminants has been made to these maps.  The hole
  in the nucleus is due to severe stellar crowding.  A few highly
  saturated stars also appear as smaller holes.  The horizontal white
  stripes at $\eta = \pm 0.2$ are due to low confidence pixels, which
  we exclude from the star count profiles.  }
\label{fig:2dmaps}
\end{figure*}

In Fig.~\ref{fig:2dmaps}, we plot the two-dimensional spatial
distribution of sources in the CMD selection boxes.  Going clockwise
from top left, the maps show RGB, AGB, RSG, and MS+BHeB point sources.
No correction for contaminants has been made.  The ellipses correspond
to $R_{dp} = 10 - 60$ kpc in steps of 10 kpc ($\rm 1\ kpc \approx
1.1\arcmin$).  The hole in the nucleus is due to severe stellar
crowding.  A few highly saturated stars also appear as smaller holes.
The horizontal white stripes at $\eta = \pm 0.2$ are due to low
confidence pixels, which have a low effective exposure time because of
dithering.  We exclude these pixels from the star count profiles
below.

In contrast to our findings for M81 and comparably deep images of
NGC~891 \citep{Mouhcine10}, there are no obvious substructures in the
distribution of MS+BHeB and RSG stars beyond the optical disc.  The
RGB and AGB stellar distributions are relatively smooth suggesting
that NGC~2403 has evolved quiescently with no significant recent major
accretions or mergers.  However, we note that the photometry in our
survey of NGC 2403 does not go as deep as in the INT or CFHT surveys
of M31 and M33 \citep{Ferguson02, Mcconnachie10}, so the very faint
structures seen in those surveys would be below our detection limit.

\subsection{Radial Profiles}
\label{sec:sdprof}

\begin{figure}
\includegraphics[width=3in,keepaspectratio=true]{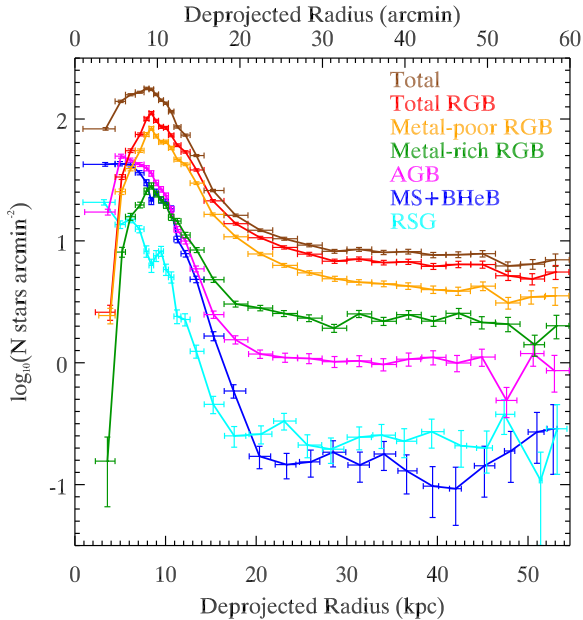}
\caption{ Raw star count profiles of point sources in the CMD
  selection boxes.  No completeness correction or background
  subtraction has been applied.  Vertical error bars include Poisson
  noise.  Horizontal error bars span the full radial range of stars in
  each bin.  Severe crowding causes the profiles to turn over near the
  nucleus at $R_{dp} \sim 5 - 10$ kpc.  Beyond 10 kpc, there are two
  regimes visible, one that extends out to $\sim 18$ kpc where the
  profiles have a steep slope and another beyond 18 kpc where they are
  relatively flat.  The metal-poor RGB profile decreases out to $\sim
  40$ kpc.  }
\label{fig:rawsdprof}
\end{figure}

Fig.~\ref{fig:rawsdprof} shows the raw surface density profiles for
point sources in the CMD selection boxes (i.e., before any
completeness correction or background subtraction has been applied).
Low confidence pixels near the chip edges and mosaic corners are
excluded from the profiles.  The lines are colour-coded so that the
total RGB box is red, metal-poor RGB box is orange, metal-rich RGB box
is green, AGB box is magenta, MS+BHeB box is blue, and RSG box is
cyan.  The top profile is the total of all the boxes.  Each point in
the profiles is the mean $R_{dp}$ of all stars within a bin.
Horizontal error bars span the full radial range of stars in each bin.
Vertical error bars include Poisson noise, which may underestimate the
true error because it does not include background galaxy clustering.
Severe crowding causes the profiles to turn over at $R_{dp} \sim 5 -
10$ kpc.  Beyond 10 kpc, there are two regimes visible, one that
extends out to $\sim 18$ kpc where the profiles have a steep slope and
another beyond 18 kpc where they are relatively flat.  In particular,
the metal-poor RGB profile decreases slowly out to $\sim 40$ kpc
showing the first indication of an outer structure of metal-poor RGB
stars around NGC~2403.

\begin{figure}
\includegraphics[width=3in,keepaspectratio=true]{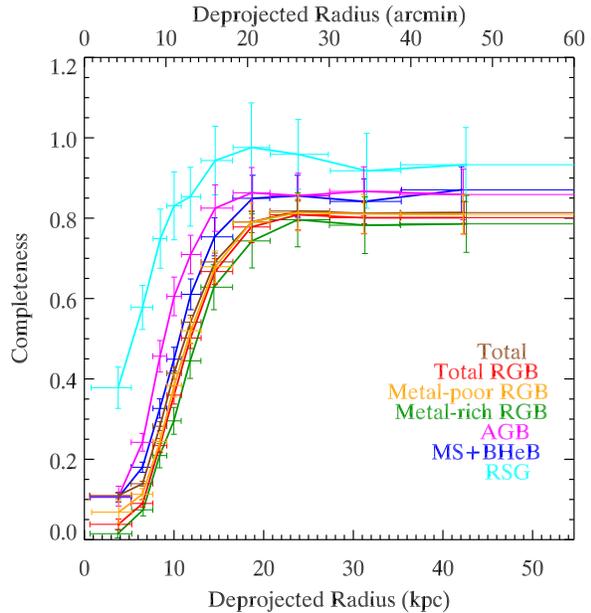}
\caption{ Completeness rate as a function of $R_{dp}$ for the CMD
  selection boxes.  Beyond 18 kpc, the completeness is approximately
  constant with radius.  Inside 18 kpc, the completeness drops with
  decreasing radius because of higher stellar crowding closer to the
  nucleus.  The total profile is $\sim 50\%$ complete at 11 kpc.  }
\label{fig:boxcomp}
\end{figure}

Next, we correct the raw star counts for completeness by weighting
each star by $w_j = 1/c_j$, where $c_j$ is the star's completeness
interpolated in colour, magnitude, and $R_{dp}$ using the artificial
stars in the appropriate field.  Fig.~ \ref{fig:boxcomp} shows the
weighted mean completeness rate (i.e., $\Sigma w_j c_j / \Sigma w_j$)
of the CMD selection boxes in the total catalogue.  Beyond 18 kpc, the
completeness is approximately constant and inside 18 kpc, the
completeness drops with decreasing radius because of stellar crowding.
This figure shows that the metal-poor RGB has a completeness rate $>
50\%$ for $R_{dp} \gtrsim 12$ kpc.  Similarly, the total profile is $>
50\%$ complete for $R_{dp} \gtrsim 11$ kpc.  Because of the difference
in seeing between the two fields, the total profile is $> 50\%$
complete for $R_{dp} \gtrsim 9$ (12) kpc in field F1 (F2) alone.

\begin{figure}
\includegraphics[width=3in,keepaspectratio=true]{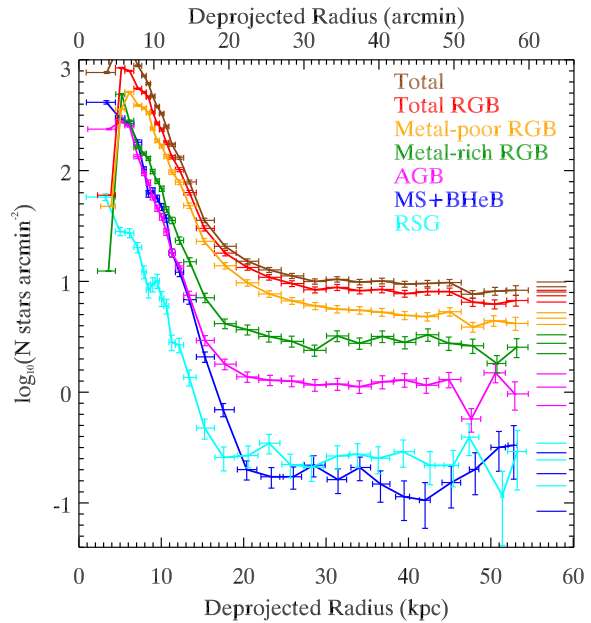}
\caption{ Completeness-corrected star count profiles.  Vertical error
  bars include Poisson noise and completeness uncertainty.  The
  background level and $1\sigma$ uncertainty for each profile,
  estimated from the last 5 bins, are marked as short horizontal lines
  at right.  The RSG, MS+BHeB, and AGB profiles quickly reach the
  background levels at $R_{dp} \sim 20$ kpc.  The metal-poor RGB
  profile exhibits a slower decline to the background level compared
  to the other profiles.  }
\label{fig:compsdprof}
\end{figure}

Fig.~ \ref{fig:compsdprof} shows the resulting completeness-corrected
surface density profiles.  The completeness correction leaves the
profiles outside 18 kpc mostly unchanged, but it has a larger effect
on the inner 18 kpc where the profile slopes are now much more similar
and more clearly show the exponential decline of the inner disc
(discussed further in \S \ref{sec:sbprof}).  Inside 6 kpc, some
profiles still turn over indicating the completeness correction is
less reliable there, which is not surprising given the high degree of
crowding.

\begin{figure}
\includegraphics[width=3in,keepaspectratio=true]{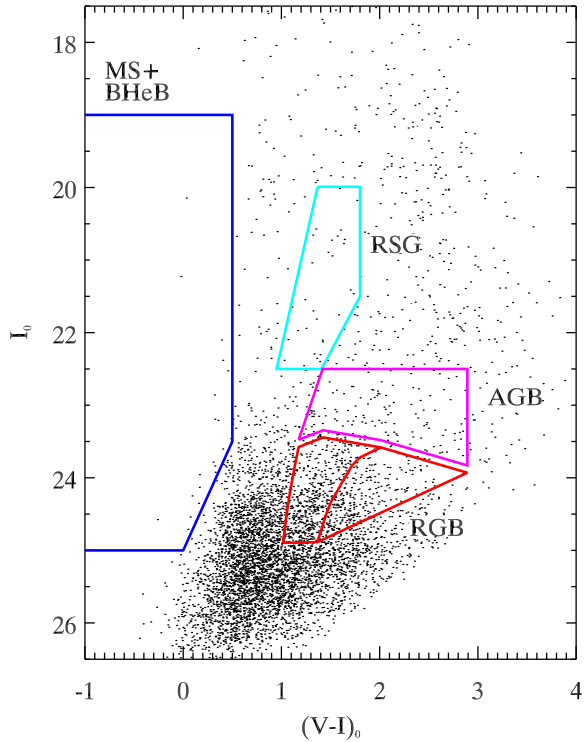}
\caption{ Background CMD covering $R_{dp} > 40.9$ kpc and an area
  $\sim 7\%$ the total field of view.  These objects are mostly
  unresolved background galaxies and MW foreground stars.  The
  selection boxes avoid the most heavily contaminated regions at
  $(V-I)_0 \sim 0 - 1$ or $I \gtrsim 25$.  The number counts of
  objects in the boxes are used to subtract contaminants from the star
  count profiles.  }
\label{fig:cmdbg}
\end{figure}

The short horizontal lines on the right-hand side of
Fig.~\ref{fig:compsdprof} mark the background levels and $1\sigma$
uncertainties.  The background level for each box comes from the total
counts and area summed over the last 5 bins covering $R_{dp} > 40.9$
kpc while the uncertainty comes from the standard deviation of the
same bins.  Fig.~\ref{fig:cmdbg} shows the CMD for the background
region, which covers an area of $\sim 130\ \rm arcmin^2$ or $\sim 7\%$
the total field of view.  There are no obvious NGC 2403 stellar
sequences visible, consistent with these objects being dominated by
unresolved background galaxies and MW foreground stars.  The selection
boxes avoid the most heavily contaminated regions at $(V-I)_0 \sim 0 -
1$ and $I_0 \gtrsim 25$.

\begin{figure}
\includegraphics[width=3in,keepaspectratio=true]{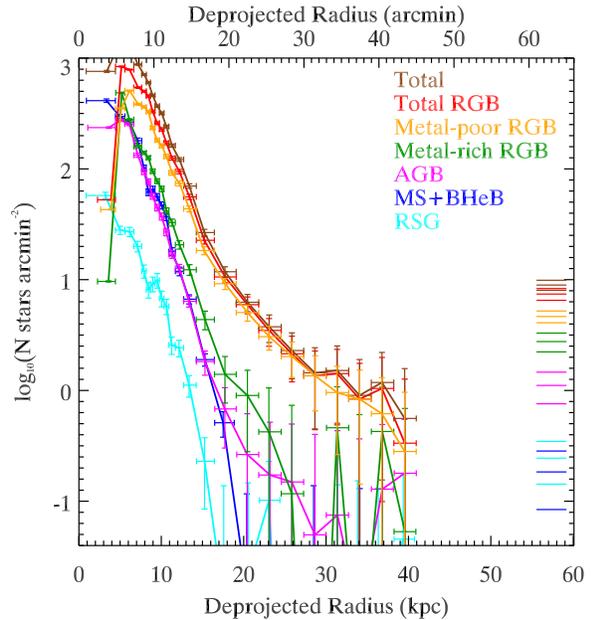}
\caption{ Background-subtracted, completeness-corrected star count
  profiles for the CMD selection boxes.  The background levels and
  $1\sigma$ uncertainties are marked by the short horizontal lines at
  right.  There is an excess of metal-poor RGB stars relative to the
  extrapolation of the inner profile outside the main disc, indicating
  the presence of a metal-poor extended structure.  The errors bars
  include the uncertainty in the background level.}
\label{fig:corrsdprof}
\end{figure}

The background-subtracted, completeness-corrected star count profiles
are displayed in Fig.~\ref{fig:corrsdprof}.  This figure shows strong
evidence for an extended metal-poor RGB component with a shallower
radial profile than the inner disc, that starts to dominate the star
counts at $R_{dp} \sim 18$ kpc.  There is also a small excess of
metal-rich RGB stars at $18 - 27$ kpc, but this could be due to
photometric errors scattering some metal-poor RGB stars into the
metal-rich box.

The MS+BHeB and RSG profiles extend out to $\sim 18$ kpc or $1.8
R_{25}$ and the AGB profile extends out to $R_{dp} = 27$ kpc or $2.8
R_{25}$, consistent with the findings of \citet{Davidge03,Davidge07}.
The AGB profile shows a slight change in slope at 18 kpc similar to
the RGB profiles, but the $1\sigma$ errors on the points beyond this
distance are too large to say with high confidence whether or not this
change is real.

All the background-subtracted profiles exhibit a similar steep slope
at $R_{dp} \sim 9 - 17$ kpc.  As we will see in Section
\ref{sec:sbprof}, this radial range is clearly dominated by outer
disc.  We fitted exponentials to the profiles in this region, after
multiplying the logarithm of the surface density by 2.5 to bring them
onto a magnitude scale, and the resulting scale-lengths are listed in
Table 2.  The quoted errors give the interval over which
$\chi_{\nu}^2$ increases by 1.0.  The young star profiles have
scale-lengths around 1.7 kpc while the AGB stars have a larger
scale-length of $\sim1.9$~kpc, suggesting some size evolution in the
scale-length over the last few Gyr.  The metal-poor RGB profile has a
longer scale-length than the metal-rich RGB profile, and both RGB
profiles have longer scale-lengths than any of the young star
profiles.  This could be due to a contribution from a distinct, more
extended metal-poor component and/or the disc RGB population being
more broadly distributed in radius than the young stars, further
suggesting outside-in formation.

\begin{table}
 \centering
 \begin{minipage}{3in}
   \caption{Radial scale-lengths for star count profiles in the range
     $R_{dp} = 9 - 17$ kpc.}
  \begin{tabular}{ll}
  \hline
   CMD box     &  Scale-length \\
               &  (kpc) \\
 \hline
MS+BHeB   & $1.70 \pm {0.17}$ \\
RSG       & $1.68 \pm { 0.49}$ \\
AGB       & $1.91 \pm { 0.24}$ \\
metal-poor RGB    & $2.56 \pm { 0.22}$ \\
metal-rich RGB    & $2.04 \pm { 0.25}$ \\
total-RGB & $2.42 \pm {0.17}$ \\
total     & $2.23 \pm { 0.13}$ \\
\hline
\end{tabular}
\end{minipage}
\label{tab:scalelengths}
\end{table}

\subsection{RGB Metallicity}
\label{sec:mdf}

\begin{figure}
\includegraphics[width=3in,keepaspectratio=true]{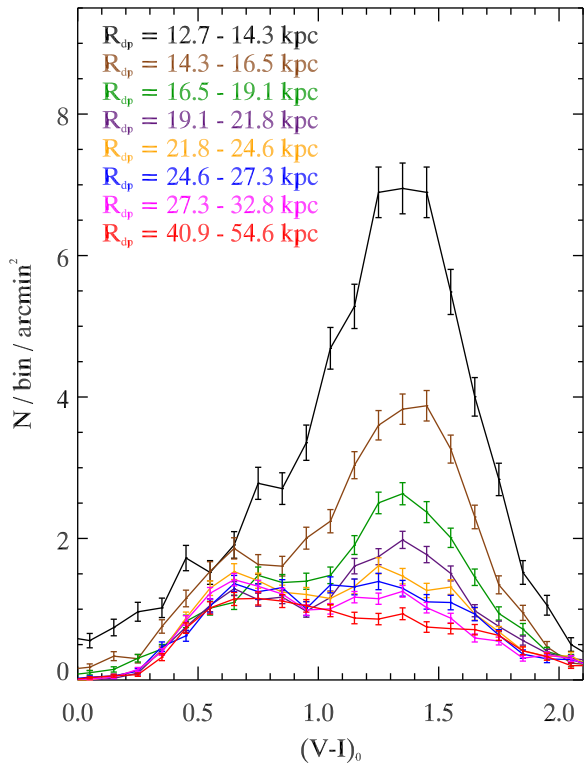}
\caption{Colour histogram for all point sources with $I_0 = 23.4 -
  24.9$ and for several different radial ranges as indicated at
  top-left.  The lowest histogram is the background region.  Error
  bars show the Poisson noise.  The peak at $(V-I)_0 = 1.3 - 1.4$
  corresponds to the RGB and is visible well past 18 kpc.  }
\label{fig:rgbchist}
\end{figure}

To examine in more detail the nature of the extended RGB component at
large radii, Fig.~\ref{fig:rgbchist} shows the colour histogram for
point sources in the magnitude range $I_0 = 23.4 - 24.9$ and for
several different radial ranges including the background region at
$R_{dp} = 40.9 - 54.6$ kpc.  Each histogram has units of surface
density per 0.1 mag-wide bin, so the background region has the lowest
values.  The error bars include the contribution from Poisson noise,
but not from clustering of background galaxies, and, therefore, they
may underestimate the true standard deviation in each bin.  In this
figure, the RGB appears as the prominent peak at $(V-I)_0 \sim 1.3 -
1.4$, which drops to the background level at the outermost radii.
There is little apparent change in the position of the peak with
radius suggesting little variation in the peak metallicity.  At
$R_{dp} = 27.3 - 32.8$ kpc, there is still a peak visible at the same
colour as the inner radii.

\begin{figure}
\includegraphics[width=3in,keepaspectratio=true]{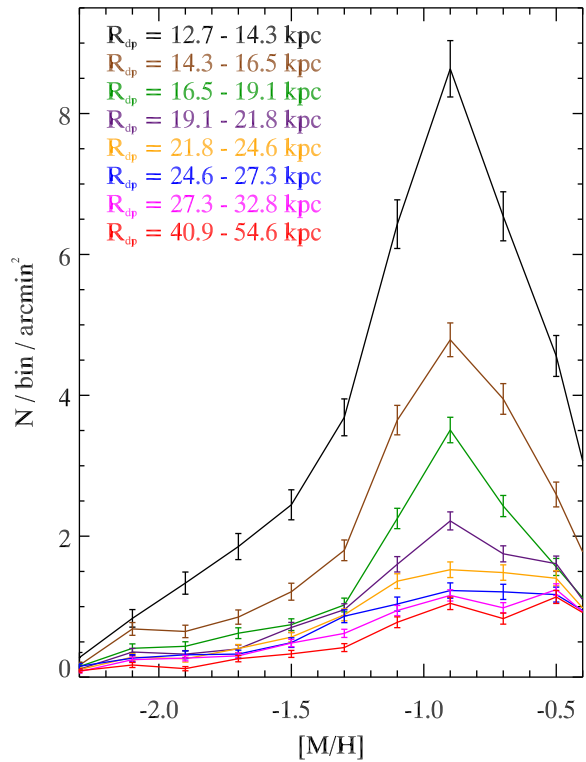}
\caption{RGB MDF for several different radial ranges using the
  \citet{Marigo08} isochrones and assuming an age of 10 Gyr and
  $[\alpha/Fe] = 0$.  Using the \citet{Dotter07a} isochrones shifts
  the peak $\sim 0.2$ dex lower.  Error bars show the Poisson noise.}
\label{fig:mdf}
\end{figure}

Next, we construct the metallicity distribution function (MDF) of
point sources falling in the RGB selection box.  For this purpose, we
assume an age of 10 Gyr and $[\alpha/Fe] = 0$ and adopt the
\citet{Marigo08} isochrones.  We use 9 isochrones spaced roughly 0.3
dex apart in the range [M/H] $=$ --2.3 to 0.2.  These isochrones form
an irregular grid in colour, magnitude, and metallicity.  We then
interpolate between the grid points to measure the metallicity of each
source in the RGB selection box.  The interpolation is performed with
the TRIGRID function in the Interactive Data Language (IDL), which
utilizes Delaunay triangulation and polynomial interpolation.

\begin{figure*}
\includegraphics[width=6in,keepaspectratio=true]{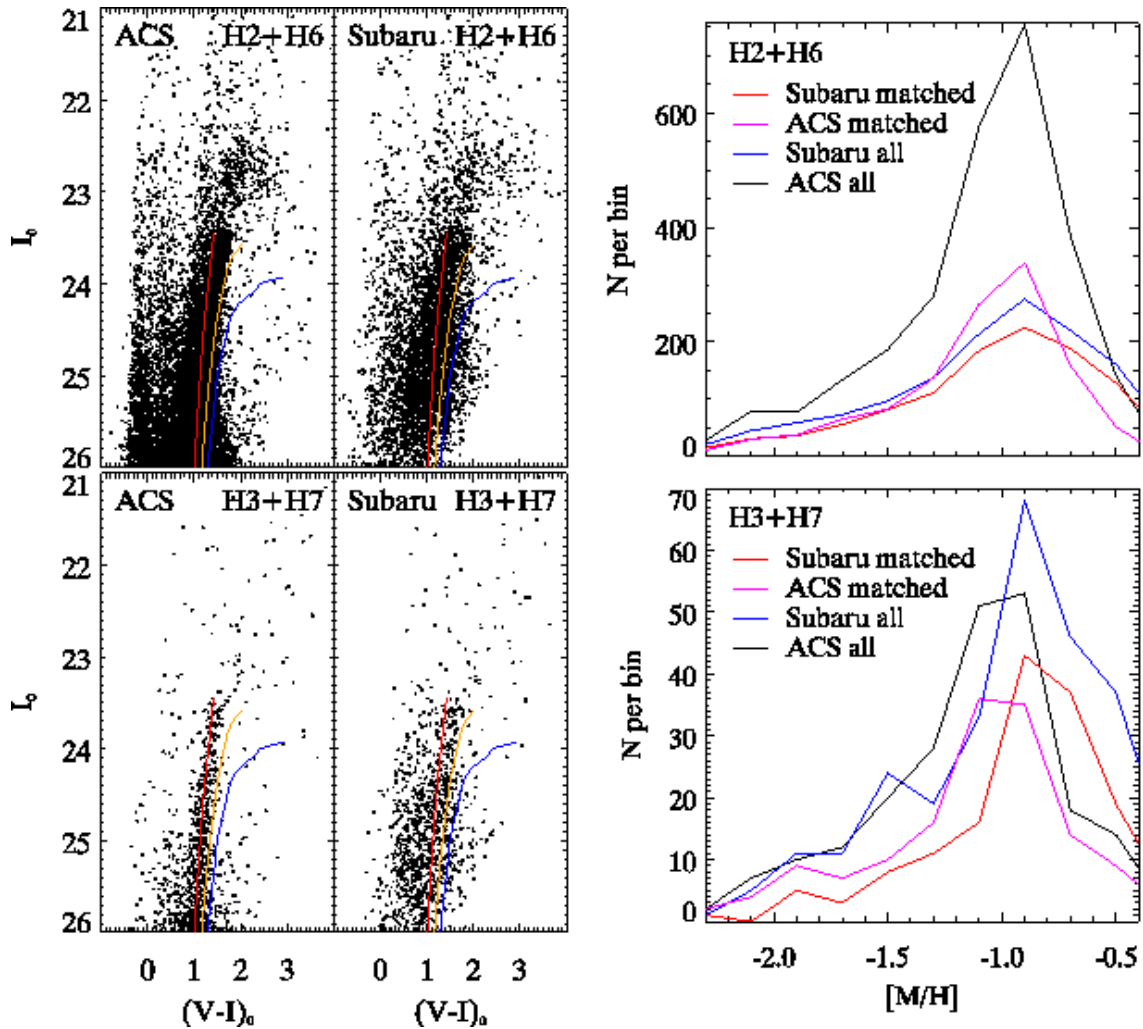}
\caption{ Comparing ACS data to Subaru data for fields H2+H6 ($R_{dp}
  \sim 11$ and 12 kpc) and H3+H7 ($R_{dp} \sim 16$ and 21 kpc).  The
  left column compares the CMDs for all point sources in the ACS and
  Subaru catalogues and \citet{Marigo08} isochrones for an age of 10
  Gyr and [M/H] $=$ --1.3, --0.7, and --0.4.  The right column shows
  the MDFs for all RGB stars and for the matched subsets in the two
  photometric metallicity systems.  The MDF peaks in the two systems
  agree to within $\sim 0.2$ dex.  }
\label{fig:acscmd}
\end{figure*}

Fig.~\ref{fig:mdf} shows the MDFs for different radial ranges in units
of surface density per 0.2 dex-wide bin.  Sources falling in the
metal-rich RGB box lie within the 2 highest metallicity bins.  We do
not show metallicities higher than --0.4 as this is roughly the
maximum metallicity contained within the metal-rich RGB box for this
age.  The MDFs are not corrected for incompleteness, but this should
not dramatically alter their shapes since the metal-rich RGB box is
typically only $\sim 5\%$ less complete than the metal-poor RGB box.

The MDFs for all radii exhibit a peak at [M/H] $= -0.9$.  Adopting the
Dartmouth isochrones \citep{Dotter07a} gives a peak metallicity $\sim
0.2$ dex lower.  Adding the above uncertainties in quadrature, and
taking the mean of the two estimates with the different isochrones,
gives [M/H] $= -1.0 \pm 0.3$ as our final estimate for the peak
metallicity.  If the mean RGB age is as young as 2 Gyr, then the peak
metallicity would be $\sim 0.4$ dex higher.  An enhancement in the
$\alpha$-elements would leave our estimate for [M/H] unchanged because
$\alpha$-enhanced isochrones can be approximated by scaled-solar
isochrones with the same global metallicity.  Using the formalism of
\citet{Salaris93}, an enhancement in the $\alpha$-elements of
$[\alpha/Fe] = 0.3$ would mean that [Fe/H] is lower than [M/H] by
$\sim 0.2$ dex.

There is no statistically significant gradient in the peak metallicity
out to 27 kpc in Fig.~\ref{fig:mdf}, but radial star formation history
variations may hide any metallicity gradient that is present given our
assumption of a simple stellar population.  As we move to larger radii
and the number of NGC~2403 stars decreases, the MDFs look increasingly
similar to the background.  At all radii, there appears to be a large
spread in metallicity.  The artificial star tests indicate that only
at $R_{dp} \lesssim 18$ kpc do the photometric errors contribute
significantly to this spread.  At these radii, there could also be
some young, red helium-burning giants contaminating the metal-poor
tail.  A spread in age at any radius may also contribute to the
observed MDF widths.  Finally, we note that a star-by-star scatter
plot of [M/H] versus $R_{dp}$ did not reveal any further information.

\citet{Davidge03} observed a single $5.5\arcmin \times 5.5\arcmin$
field located on the NE minor axis at $R_{dp} \approx 18.5$ kpc.  He
measured an RGB metallicity gradient by comparing the observed RGB
colour near the RGB tip to that of several MW globular clusters.  He
divided his field into radial subregions and found $\rm [Fe/H] = -0.8
\pm 0.1\ (ran) \pm 0.3\ (sys)$ at $R_{dp} \sim 15$ kpc and $\rm [Fe/H]
= -2.2 \pm 0.2\ (ran) \pm 0.8\ (sys)$ at $R_{dp} \sim 22$ kpc (after
converting his galactocentric radii to reflect our values of
inclination, position angle, and distance modulus).  We have extracted
the RGB stars within his observed field and divided them into the same
subregions, but the MDFs were consistent with those in
Fig.~\ref{fig:mdf} and they showed no evidence for a metallicity
gradient.  We note that Davidge's field encompasses a radial range
where the stellar density falls off rapidly; it is possible that an
underestimate of the background contamination could explain his
result.

To check the external accuracy of our metallicities, we downloaded and
reduced four HST/ACS fields from the MAST archive (PID 10523) that
fall within our surveyed area.  The names and locations of the fields
are given in Fig.~\ref{fig:fields}.  We performed the data reduction
using the ACS module of the DOLPHOT package~\footnote{DOLPHOT is an
  adaptation of the photometry package HSTphot \citep{Dolphin00}. It
  can be downloaded from http://purcell.as.arizona.edu/dolphot/.}
following the steps outlined in the DOLPHOT manual and using the
default input parameters.  We defined objects as point sources if they
were classified by DOLPHOT as 'good stars' with $S/N > 5$ and crowding
parameter $< 0.5$ in both filters, and if the overall $\vert
sharp\vert < 0.1$ and $\chi < 3$.  Magnitudes were reported for every
source in the native ACS VEGAMAG filter system ($F606W$ and $F814W$)
and in the ground-based Johnson-Cousins system using the
transformation equations in \citet{Sirianni05}.

Fig.~\ref{fig:acscmd} compares the ACS CMDs and MDFs with the Subaru
CMDs and MDFs for all point sources falling within the ACS fields.  To
boost the number statistics, we combined the ACS fields into two pairs
with similar radii, $R_{dp} \sim 11$ and 12 kpc for H2$+$H6 and
$R_{dp} \sim 16$ and 21 kpc for H3$+$H7.  Overplotted on the CMDs are
the \citet{Marigo08} isochrones for an age of 10 Gyr and [M/H] $=$
--1.3, --0.7, and --0.4.  For ease of comparison, the ACS CMDs are in
the ground-based magnitude system, but to avoid any possible biases
introduced in the transformation, the computation of the ACS MDFs was
done in the native ACS system.  Additionally, we matched point sources
in the Subaru catalogue with point sources in the ACS catalogues by
applying constant offsets of $\sim 0.2\arcsec - 1.4\arcsec$ in right
ascension and declination.  The MDFs for the matched subsets are also
shown in Fig.~\ref{fig:acscmd}.

Overall, there is good correspondence between the ACS and Subaru CMDs.
The RGB is clearly visible in both CMDs despite the higher background
contamination in the ground-based data which dominates in H3$+$H7 at
$0 \lesssim (V-I)_0 \lesssim 1$.  In H2+H6, the Subaru CMD is missing
many of the blue sources at $(V-I)_0 \sim 0$ because they lie in
highly crowded young stellar associations.  The Subaru MDFs have
broadly similar shapes as the ACS MDFs, with the caveat that the
Subaru completeness is lower in H2+H6 than in H3+H7 because of the
higher crowding level.  This experiment indicates that our peak
metallicities agree with ACS to within 0.2 dex.

\section{Quantifying The Global Structure of NGC 2403}

\label{sec:sbprof}

\begin{figure}
\includegraphics[width=3in,keepaspectratio=true]{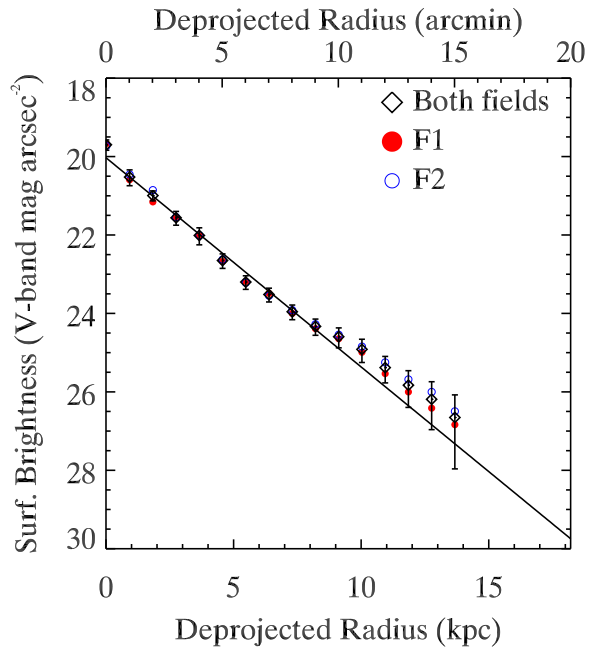}
\caption{ Diffuse light SB profile for F1 (solid circles) F2 (open
  circles), and the average of both fields (diamonds).  Error bars are
  shown only for the average profile and they include read noise, sky
  background uncertainty, and the r.m.s. variation within each
  annulus.  The best-fit exponential disc (solid line) has a
  scale-length $h = 2.0 \pm 0.2$ kpc.  }
\label{fig:diffuselight}
\end{figure}

Having established the existence of an extended structural component
at large radii in NGC~2403, we now construct a composite SB profile
which uses diffuse light and resolved star counts in the regions where
they are each most reliable.  Within the bright optical disc, where
the effects of incompleteness on the star counts are most severe, we
use the diffuse light because of its insensitivity to these effects.
In the outer regions, where the sky background dominates the diffuse
light, but where the completeness rate is the highest and varies the
least, we use the total star counts, which have a higher contrast over
the background than the diffuse light.

We derived the V-band diffuse light SB profile of NGC~2403 using the
IRAF {\it ellipse} task with elliptical annuli of constant center,
position angle, and inclination after masking saturated stars.  In
each elliptical annulus, the median pixel value was computed after two
$5\sigma$ clipping iterations.  The diamonds in
Fig.~\ref{fig:diffuselight} show the sky-subtracted profile for the
average of both fields while the circles show the profile for each
field separately.  The sky value for each field was estimated from the
mode of the pixel histogram, which was 1930 ADU and 2660 ADU in F1 and
F2, respectively.  These values translate to a sky SB of $\mu_V =
21.9\ \magsec$ and $\mu_V = 21.6\ \magsec$.  Another method of sky
estimation, that involved taking the mean of the median pixel value in
16 3\arcmin x3 \arcmin boxes near the edges of the mosaic, gave values
10 ADU and 18 ADU less than the first method, or about 0.9 and 1.5
times the standard deviation of the median pixel box values in F1 and
F2, respectively.  To be conservative, we adopt the difference between
the two methods as the sky uncertainty.  The error bars are shown only
for the average profile and they include read noise, sky background
uncertainty, and the root-mean-square (r.m.s.) deviation of the pixel
values in each annulus to account for azimuthal variations due to
spiral arms, HII regions, OB associations, and any possible warping of
the stellar disc.  The median foreground extinction of all point
sources, $A_V = 0.12$ mag, was subtracted from the profiles, but no
correction was made for internal extinction.

The best-fit exponential disc (solid line) has a scale-length of $h =
2.0 \pm 0.2$ kpc, in agreement with the star count profiles in Table
2.  The diffuse light profile shows a change to a slightly flatter
slope at $\sim 7$ kpc., which is more pronounced in F2. Inside 4.5
kpc, the scale-length is $h = 1.7 \pm 0.2$ kpc, which is in good
agreement with the value of $h = 1.5$ kpc derived by \citet{Okamura77}
over the same radial range.

\begin{figure}
\includegraphics[width=3in,keepaspectratio=true]{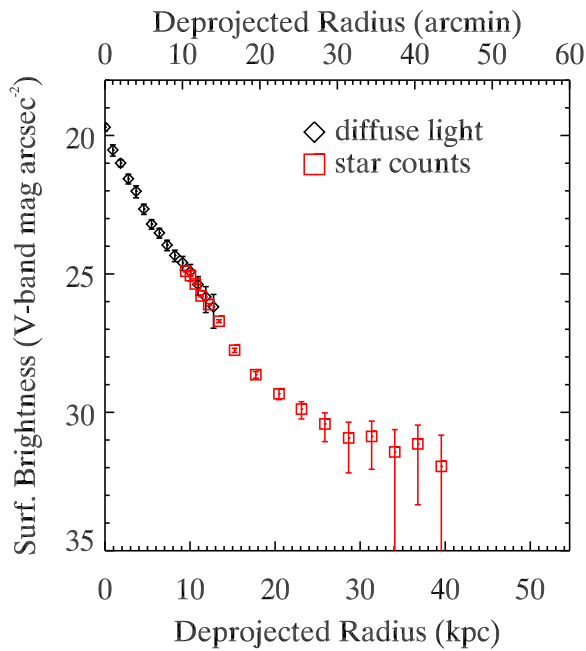}
\caption{Composite SB profile for NGC~2403 made from the diffuse light
  (diamonds) and total star count profiles (squares).  The composite
  profile traces the SB of NGC~2403 down to $\mu_V \sim 32\ \magsec$.}
\label{fig:mergedprof}
\end{figure}

The next step in constructing the SB profile involves merging the star
count profile with the diffuse light profile.  The conversion of star
counts to magnitudes is achieved by the relation $\mu(R) = -2.5\rm\
log_{10}[\Sigma(R)]+ZP$, where $\Sigma(R)$ is the stellar surface
density and the zero-point, ZP = 31.32, is estimated from the
overlapping region between the star counts and diffuse light.  By
merging the profiles in this way, we can trace NGC~2403's SB over a
larger radial range than is possible with either profile alone.  This
is made possible by the much lower sky background attained with the
resolved star counts, $\mu_V \sim 29\ \magsec$, more than 7 magnitudes
fainter than the diffuse light background.

\begin{figure}
\includegraphics[width=3in,keepaspectratio=true]{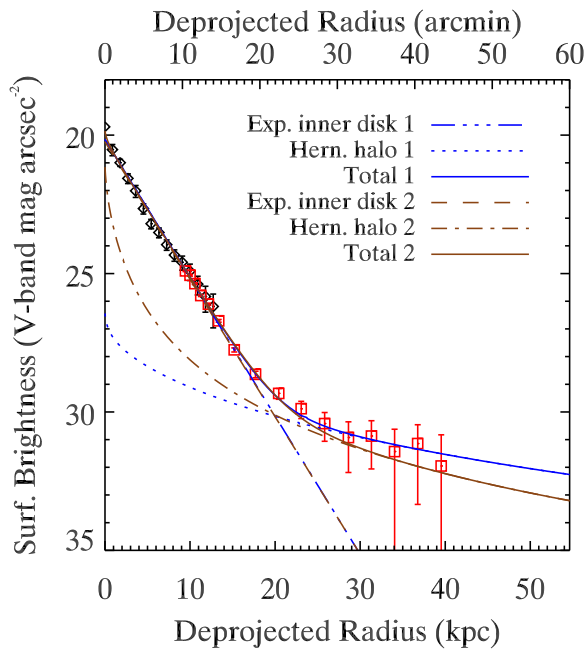}
\caption{Two different disc+halo decompositions of the SB profile: (1)
  exponential disc with scale-length $h = 2.17 \pm 0.03$ kpc and
  Hernquist halo with scale-radius fixed at the MW's value $r_s =
  14.0$ kpc, (2) exponential disc with $h = 2.18 \pm 0.03$ kpc and
  Hernquist halo with $r_s = 1.3^{+1.5}_{-0.5}$ kpc.  The points and
  error bars are the same as in Fig.~\ref{fig:mergedprof}.  Solid
  lines show the total model profiles and broken lines show individual
  components.  }
\label{fig:hern}
\end{figure}

In Fig.~\ref{fig:mergedprof}, we show the composite SB profile for
NGC~2403 made by combining the diffuse light (diamonds) and total star
count profiles (squares).  The SB profile extends over 12 magnitudes
and reaches $\mu_V \sim 31\ \magsec$ at $R_{dp} \sim 30$ kpc and
$\mu_V \sim 32\ \magsec$ at $R_{dp} \sim 40$ kpc.  Disregarding any
contribution from the inner disc beyond 22 kpc, the extended component
can be described by an exponential scale-length $h = 8.7 \pm 1.1$ kpc
or a projected power-law ($I(r) \propto r^{-\gamma}$) with index
$\gamma = 3.4 \pm 0.1$.

Merging the diffuse light and star counts assumes that any radial
gradient in the colour and luminosity function has a negligible effect
on the SB.  However, we are primarily concerned with the outskirts of
NGC~2403, where the luminosity function is dominated by the RGB and
varies relatively little with radius.  One other related subtlety is
that the star count profiles should be weighted by the $V$-band
luminosities of their constituent stars before summing them to make
the total profile.  We can estimate the magnitude of this effect by
scaling the star count profiles by the average V-band luminosities of
point sources in their respective CMD boxes.  This has the effect of
shifting the young star and AGB profiles upward relative to the RGB
profiles and slightly steepening the total star count profile
scale-length in the region $R_{dp} = 9 - 17$ kpc from 2.2 kpc to 2.0
kpc.  Thus, we expect that disregarding this effect will have little
impact on our results.  We also note that correcting for internal
extinction in the main gas disc could steepen the profile, as well,
but we have chosen not to attempt any correction for this because of
the uncertain dust properties in NGC 2403.


\begin{figure}
\includegraphics[width=3in,keepaspectratio=true]{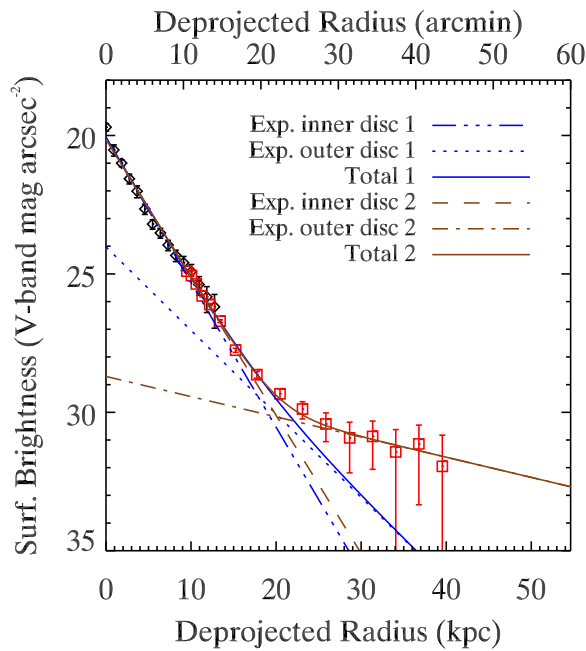}
\caption{Two different disc+disc decompositions of the SB profile: (1)
  inner disc exponential scale-length $h = 2.08 \pm 0.03$ kpc and
  outer disc scale-length held fixed at the MW thick disc's value, $h
  = 3.6$ kpc, (2) inner disc scale-length $h = 2.18 \pm 0.03$ kpc and
  outer disc scale-length $h = 15 \pm 8$ kpc.  The points and error
  bars are the same as in Fig.~\ref{fig:mergedprof}.  Solid lines
  shows the total model profiles and broken lines show individual
  components.  }
\label{fig:exp}
\end{figure}

Fig.~\ref{fig:hern} shows two separate models for NGC~2403's SB
profile which include an exponential inner disc and a
spherically-symmetric Hernquist halo.  We recall that a Hernquist
profile can be characterised by a scale radius, $r_s$, which is
approximately $41\%$ of the half-mass radius and that, for $r >> r_s$,
the projected light profile follows a power-law with exponent $\gamma
= 3$ \citep{Hernquist90}.  In the first model, the best-fit
exponential inner disc scale-length is $h = 2.17 \pm 0.03$ kpc and the
halo's scale radius is held fixed at $r_s = 14.0$ kpc.  We use 14.0
kpc as a fiducial value because it was recently estimated for the MW's
halo \citep{Newberg06}, it lies within the range of $10 - 20$ kpc
found in some theoretical semi-analytic simulations \citep{Bullock05},
and it provides a reasonable description of M81's extended component
\citep{Barker09}.  If we allow the halo's scale radius to be free,
then its best-fit value is $r_s = 1.3^{+1.5}_{-0.5}$ kpc and the disc
scale-length is virtually unchanged at $h = 2.18 \pm 0.03$ kpc.  Both
values of $r_s$ provide similar fit qualities and yield similar halo
luminosities.  If we extrapolate the fits out to 50 kpc, then the
haloes would contain $\sim 1 - 5\%$ of the total galactic V-band
luminosity, or $L_V \sim 1 - 5 \times 10^8 \L_{\sun}$.  Note that the
luminosity does not significantly change if extrapolated further out
to 100 kpc.

If the extended component is a disc structure, then it would be more
appropriate to describe it with a radial exponential profile than a
Hernquist profile.  We show two different disc plus disc models of the
SB profile in Fig.~\ref{fig:exp}.  In the first, the scale-length of
the extended component is fixed to that of the MW's thick disc, $h =
3.6$ kpc \citep{Juric08} and the resulting best-fit scale-length of
the inner disc is $h = 2.08 \pm 0.03$ kpc.  However inspection of
Fig.~\ref{fig:exp} shows this model clearly fails to explain the
excess light beyond 20~kpc. Thus, the scale-length of the extended
component could only be as small as the MW's thick disc if we have
grossly underestimated the background or if there is a third
structural component that dominates at $R_{dp} \gtrsim 30$ kpc.  In
the second model, the extended component's scale-length is free and
its best-fit value is $h = 15 \pm 8$ kpc while the best-fit inner disc
scale-length is $h = 2.18 \pm 0.03$ kpc.  The V-band luminosity of the
outer disc in these two models is $\sim 1 - 7\%$ of the total.

We conclude from these decompositions that the data can accomodate a
wide range of scale radius or scale-length for the extended component
given the signal-to-noise ratio and limited radial range of the outer
points in the SB profile.  Both power-law and exponential fits can be
found which provide acceptable fits. The extended component's $V$-band
luminosity is somewhat dependent on the model adopted, particularly on
the behaviour of the model inside 20 kpc.  If this component exists at
all radii, then the range $\sim 1 - 7 \times 10^8\ L_{\sun}$ ($\sim 1
- 7\%$ of the total) is likely to bracket the true luminosity.

\section{Discussion}
\label{sec:disc}

We have found strong evidence for an extended structure of RGB stars
in NGC~2403 which dominates the light profile beyond $R_{dp} \sim 18$
kpc and has a peak metallicity of [M/H] $= -1.0 \pm 0.3$.  This
structure has a flatter radial profile than the inner disc, can be
reliably traced to $R_{dp} \sim 40$ kpc and $\mu_V \sim 32\ \magsec$.
The radial profile is consistent with a power-law halo or exponential
disc and we now discuss these possibilities in more detail.

The MW and M31 have the most thoroughly studied stellar haloes, so
they provide a useful baseline for comparison to the extended
component in NGC 2403.  Although the MW's stellar halo contains
significant amounts of substructure \citep[e.g.][]{Bell08}, many
studies over the years have found that it can be broadly described by
a power-law volume density distribution with index $\Gamma = \gamma +
1 \sim$ 3.0 (see the review by \citet{Helmi08}).  \citet{Newberg06}
described the spatial distribution of halo stars with a Hernquist
profile with scale radius $r_s \approx 14$ kpc.  The total luminosity
of the MW halo was estimated by \citet{Carney90} to be $L_V \sim 10^9
L_{\sun}$, which is $\sim 5\%$ of the total galactic luminosity of
$\sim 2 \times 10^{10}\ L_{\sun}$ \citep{Sackett97}.

Accretion remnants are also observed throughout the halo of M31, but
\citet{Ibata07} found that a subregion of the southern quadrant lacked
any substructure and could be fit out to 150 kpc by a Hernquist
profile with $r_s = 53.1 \pm 3.5$ kpc or an exponential profile with
scale-length $h = 46.8 \pm 5.6$ kpc.  Fitting the minor axis profile
in regions devoid of spatial substructures, they derived a projected
power-law index of $\gamma = 1.91 \pm 0.12$.  They estimated a total
halo $V$-band luminosity similar to the MW's halo, corresponding to
$\sim 2\%$ of the total of $\sim 4 \times 10^{10}\ L_{\sun}$
\citep{deVauc91}.

In their analytic simulations of hierarchical galaxy formation,
\citet{Purcell07} found that the fraction of diffuse halo light
increases with host galaxy total mass.  The fractional luminosity of
the extended component in NGC 2403 is of the same order as the MW and
M31 stellar haloes, and higher than the expected value of $\sim 0.3\%$
for a $10^{11}\ M_{\sun}$ dark matter (DM) halo in the
\citet{Purcell07} simulations.  Their simulations had considerable
scatter at fixed host galaxy mass reflecting variations in the mass
accretion history.  Indeed, a fractional luminosity of 1\%, on the low
end of our estimates, is within the 95\% confidence interval for their
simulations of haloes with this mass.  Given the uncertainties in the
mass of NGC 2403's DM halo and the star formation prescriptions
employed in the simulations, these differences are not entirely
surprising and cannot be used to exclude the idea that the extended
component is a stellar halo.  What is more clear is that the extended
component in NGC 2403 is less luminous than the MW and M31 haloes, as
would be expected if stellar halo luminosity scales with total galaxy
luminosity.

\begin{figure}
\includegraphics[width=3in,keepaspectratio=true]{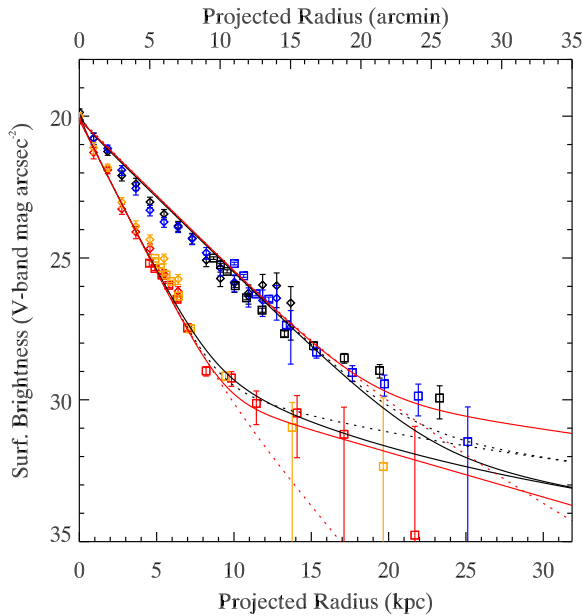}
\caption{Major and minor axis SB profiles (upper and lower points,
  respectively) constructed using circular annuli in $40\degr$-wide
  wedges centred on the axes.  The diamonds and squares show the
  diffuse light and total star count profiles, respectively.  The
  points are colour-coded so that the north-west major axis is black,
  south-east major axis is blue, north-east minor axis is red, and
  south-west minor axis is orange.  The black lines show the disc+halo
  models from Fig.\ \ref{fig:hern} with halo scale radius $r_s = 14.0$
  kpc (dotted lines) and 1.3 kpc (solid lines).  The red lines show
  the disc+disc models from Fig.\ \ref{fig:exp} with outer disc
  scale-length $h = 3.6$ kpc (dotted lines) and 15 kpc (solid lines).
  The data prefer a flattened geometry for the extended component with
  a projected axis ratio similar to that of the inner disc.  }
\label{fig:az}
\end{figure}

The MW halo has a peak metallicity [Fe/H] $\sim -1.6$, or [M/H] $\sim
-1.3$ for a typical $\rm [\alpha/Fe] \sim 0.4$ (see the reviews by
\citet{Helmi08} and \citet{Geisler07}).  Several other recent studies
have found evidence for an extended halo in M31 with metallicities in
the range --0.7 to --1.5
\citep{Reitzel02,Kalirai06,Chapman06,Richardson09}.  \citet{Font06}
and \citet{Purcell08} found that the stellar halo metallicity in
semi-analytic and analytic simulations correlated positively with
stellar halo mass.  The metallicity we measure for the extended
component in NGC~2403 is roughly similar to that in the MW and M31
haloes, but given the precision of our measurement we cannot say with
high confidence that it is inconsistent with a halo mass-metallicity
relation.  Again, it is important to remember that the theoretical
simulations predict a spread in halo properties at a given mass
depending on the exact details of the accretion history.

The concept of a single well-defined stellar halo in the MW may not be
totally accurate.  \citet{Carollo10} found evidence that the MW halo
could be divided into two kinematically distinct components, with a
flattened inner halo dominating at $\sim 5 - 10$ kpc and a roughly
spherical outer halo dominating beyond 20 kpc.  The inner halo density
distribution had a power-law index $\Gamma = 3.17 \pm 0.2$ and an MDF
that peaked at [Fe/H] $= -1.6$ whereas the outer halo had $\Gamma =
1.79 \pm 0.29$ and a peak [Fe/H] $= -2.2$.  These results highlight
the fact that the inferred properties can depend on the region
observed.  If the NGC 2403 extended component is a dual-component halo
with the same radial divisions as the MW halo, then, because the inner
disc dominates the light at small radii, our observations would be
most sensitive to the transition zone between the inner and outer
haloes and the beginning of the outer halo.

To explore further if the extended component could be a spherically
symmetric halo, Fig.~\ref{fig:az} plots the SB profile along the major
and minor axes (upper and lower points, respectively).  These profiles
were constructed using circular annuli in $40\degr$-wide wedges
centred on each axis.  The diamonds and squares show the diffuse light
and total star count profiles, respectively.  The points are
colour-coded so that the north-west major axis is black, south-east
major axis is blue, north-east minor axis is red, and south-west minor
axis is orange.  The black lines show the disc+halo models from \S
\ref{sec:sbprof} with halo scale radius $r_s = 14.0$ kpc (dotted
lines) and 1.3 kpc (solid lines).  The red lines show the disc+disc
models from \S \ref{sec:sbprof} with outer disc scale-length $h = 3.6$
kpc (dotted lines) and 15 kpc (solid lines).  To overplot the disc
models on the minor axis, the radius was divided by the axis ratio of
the inner disc.  The spherical halo models underpredict the SB on the
major axis beyond $R_{pr} \sim 17$ kpc.  The disc+disc model with
outer disc scale-length $h = 15$ kpc provides the best fit to the
major axis profiles.  Thus, the data show a clear preference for a
flattened geometry for the extended component with a projected axis
ratio similar to that of the inner disc.

If it is a disc structure, the extended component may be analagous to
the thick discs observed in the MW and edge-on galaxies.  Since we
lack a full understanding of how thick discs form, it is unclear
exactly how their stellar populations should scale with host galaxy
mass.  The scale-length and luminosity of the extended component in
NGC 2403 are longer and smaller, respectively, than those of the MW's
thick disc, which has a radial scale-length of $\sim 2 - 4$ kpc
\citep{Juric08,Carollo10}, and contributes $\sim 15\%$ to the MW's
total disc luminosity \citep{Buser99,Chen01,Larsen03}.  The extended
component's scale-length is also larger than that of the extended disc
and the thick disc in M31, which are $h = 5.1 \pm 0.1$ kpc and $h =
8.0 \pm 1.2$ kpc respectively \citep{Ibata05,Collins11}, as well as
the thick disc scale-lengths measured by \citet{Yoachim06} for edge-on
spirals with similar circular velocities.  The ratio of outer-to-inner
disc luminosities in NGC~2403 is lower than most of their sample, as
well.  The metallicity of the extended component is lower than the
mean metallicity of the MW's thick disc, [Fe/H] $\sim -0.6$
\citep{Gilmore95,Robin96,Soubiran03,AllendePrieto05}, but is similar
to the values of [Fe/H] $\sim -0.9$ and $-1.0$ found in for the
extended and thick discs in M31 \citep{Ibata05,Collins11}.  Thick
and/or extended disc structures have been hypothesized to form through
a variety of processes, including direct accretion, heating of a
pre-existing disc and radial migration \citep[e.g.][]{Penarrubia06,
  Richardson08, Schonrich09,Loebman10}.

It is not straightforward to compare in detail our results for
NGC~2403 with those found for other systems beyond the Local Group.
Most wide-field studies published to date have focused on the vertical
profiles of highly-inclined systems
\citep{Mouhcine10,Bailin11,Tanaka2011} while interpretation of the HST
and Gemini studies is signficantly complicated by the small FOV
coverage \citep[e.g.][]{Tikhonov05, Vlajic09, Rejkuba09}.
Nonetheless, evidence is mounting for the ubiquitous presence of
moderate metallicity stellar envelopes surrounding a variety of galaxy
types \citep[e.g.][]{Seth05,Mould05, Ibata09,Tanaka2011} and our
results for NGC~2403 provide further support for this.  One system for
which a detailed comparison is possible is the moderately-inclined
MW-analogue M81. This system had a close encounter with its two
neighbors, M82 and NGC 3077, some $\sim 200 - 300$ Myr ago
\citep{Yun94,Yun99}.  \citet{Barker09} identified an extended stellar
component around M81 that could be its halo or an extended disc like
that in M31.  This component bears some striking differences to the
extended component in NGC 2403.  Disregarding any contribution from
the inner disc, the total star counts beyond $R_{dp} \sim 20$ kpc had
a radial scale-length of $h = 12.9 \pm 0.9$ kpc and power-law index of
$\gamma = 2.0 \pm 0.2$.  The extended component began to dominate the
SB profile at $\mu_V \sim 26\ \magsec$ and the implied luminosity was
$L_V \sim 3 - 6 \times 10^9 L_{\sun}$ if extrapolated out to 100 kpc
over all position angles.  There was an RGB visible in the CMD at
$R_{dp} = 32 - 44$ kpc, which was estimated to have a peak metallicity
[M/H] $= -1.1 \pm 0.3$ assuming an age of 10 Gyr.  In addition, AGB
and metal-rich RGB stars were detected out to 40 kpc at surface
densities of $\sim 0.5\ \rm stars\ arcmin^{-2}$.  The azimuthal star
count profile for the metal-poor RGB was somewhat flatter than that
for the metal-rich RGB and AGB, suggesting that the extended component
was a halo or a more face-on or thicker disc structure.

In contrast, the extended component in NGC~2403 starts to dominate
over the disc at a lower SB and the implied luminosity is $\sim 10$
times less.  The radial profile is steeper and there are no detectable
AGB or metal-rich RGB stars beyond 30 kpc.  The axis ratio of the
extended component does not appear significantly different from that
of the inner disc out to $R_{pr} \sim 25$ kpc.

The few similarities that do exist between the M81 and NGC~2403
extended components include the fact that both start to dominate over
the bright optical disc at $R_{dp} \sim 20$ kpc, both have radial
profiles flatter than the inner disc, and both are dominated by RGB
stars with similar metallicities.  Also, the lack of a bulge in
NGC~2403 suggests that the presence of an extended component does not
require a bulge and that the two components may be unrelated.  Indeed,
\citet{Barker09} found that a single $r^{1/4}$-law profile could not
fit both M81's bulge and its extended component.

It is interesting, then, that NGC~2403 so far shows no clear-cut signs
of interaction in its stellar or HI distributions yet it still has an
extended component that could be a thick disc or halo.  Our survey has
found no obvious stellar streams or substructures around NGC 2403 that
would attest to a significant recent accretion, but structures like
the extended tails of RGB stars observed around M33
\citep{Mcconnachie10} would be too faint for us to observe.  Regions
beyond the area we have surveyed are more likely to contain
substructures because of the longer phase mixing timescale there
\citep{Johnston08}.

Finally, we note that the main (inner) exponential disc of NGC~2403
dominates the surface brightness profile to $\sim 18$~kpc, or $\ga 8$
scalelengths, and is characterized by a metallicity of [M/H]=--1.0 in
its outermost parts. This is reminiscent of NGC~300 where
\citet{Vlajic09} trace the exponential disc to $\sim 10$ scalelengths
(15 kpc) where it has a peak metallicity of [Fe/H] = --0.9.  This
finding provides further evidence that the stellar discs of spiral
galaxies can often be far more extended than commonly thought.


\section{Summary}
\label{sec:summ}

Using Suprime-Cam on the Subaru telescope we have conducted a
wide-field imaging survey of RGB stars around the low mass spiral
galaxy NGC~2403.  These observations represent the first global
analysis of RGB stars in a late-type spiral beyond the Local Group.
The surveyed area reaches a maximum $R_{pr} \sim 30$ kpc or $R_{dp}
\sim 60$ kpc.  The CMD reaches 1.5 mag below the tip of the metal-poor
RGB at a completeness rate $> 50\%$ for $R_{dp} \gtrsim 12$ kpc.  We
detect young stars (ages $\sim 10 - 200$ Myr) out to radii $\sim 1.8
R_{25}$, or $R_{dp} \sim 18$ kpc.

Using the combination of diffuse light photometry and resolved star
counts, we are able to trace the SB profile over a much larger range
of radius and magnitude than possible with either technique alone.
The exponential disc as traced by RGB stars dominates the SB profile
out to $\ga 8$ disc scale-lengths, or $R_{dp} \sim 18$ kpc, and
reaches a $V$-band SB of $\mu_V \sim 29\ \magsec$.  Beyond this
radius, we find strong evidence for an extended structural component
with a flatter SB profile than the inner disc and which we trace out
to $R_{dp} \sim 40$ kpc and $\mu_V \sim 32\ \magsec$.  The extended
component's V-band luminosity integrated over all radii is one to a
few percent that of the whole galaxy, depending on assumed profile.
At $R_{dp} \sim 20 - 30$ kpc, we estimate a peak metallicity [M/H]~$=
-1.0 \pm 0.3$ assuming an age of 10 Gyr and $\rm [\alpha/Fe] = 0$.
The projected axis ratio of the extended component does not appear
significantly different from that of the inner disc within $R_{pr}
\sim 25$ kpc.

Possible interpretations for the nature of this component include a
inestellar halo or thick disc.  Kinematic information for tracer
populations would help distinguish whether this component is a
rotating disc structure or a pressure-supported halo.  There are few,
if any, bright AGB stars in this component to act as spectroscopic
targets, so we must wait for the next generation of facilities to
target the more numerous RGB stars.  These results provide further
evidence that faint, extended stellar structures appear to be a
generic feature of disc galaxies, even isolated late-type systems.

\section*{Acknowledgments}

MKB and AMNF acknowledge support from a Marie Curie Excellence Grant
from the European Commission under contract MCEXT-CT-2005-025869 and a
rolling grant from the Science and Technology Facilities Council.


\bibliographystyle{mn2e}

\bibliography{references}

\bsp

\label{lastpage}


\end{document}